%% file: aliindex.tex
\documentclass[letterpaper,twocolumn,10pt]{article}
\usepackage{usenix-2020-09}

\usepackage[hang,flushmargin]{footmisc}
\usepackage[normalem]{ulem}
\usepackage{etex}
\usepackage{tikz}
\usepackage{amsmath}

\usepackage{filecontents}
\usepackage{color}
\usepackage{xcolor}
\usepackage[compact]{titlesec}
\usepackage{xspace}
\usepackage{enumitem}
\usepackage{mfirstuc}
\usepackage{cleveref}
\crefname{section}{§}{§§}
\usepackage{caption}
\usepackage[labelformat=simple]{subcaption}

\usepackage{circledsteps}
\usepackage[ruled,vlined,linesnumbered]{algorithm2e}

\usepackage{minted}
\usepackage{microtype}
\usepackage{xurl}

\usepackage{booktabs}
\usepackage{multirow}
\usepackage{pifont}
\usepackage{tablefootnote}
\usepackage{amssymb}
\usepackage{makecell}
\makeatletter
\renewcommand{\@cite}[2]{%
  \textcolor{black}{[}%
  \textcolor{blue}{#1}%
  \if@tempswa %
    \textcolor{black}{,}%
    \textcolor{blue}{#2}%
  \fi
  \textcolor{black}{]}%
}
\makeatother
\usepackage{bookmark}
\makeatletter
\renewcommand{\sectionautorefname}{\S\@gobble}
\renewcommand{\subsectionautorefname}{\S\@gobble}
\renewcommand{\subsubsectionautorefname}{\S\@gobble}

\setenumerate[1]{itemsep=0pt,partopsep=0pt,parsep=\parskip,topsep=5pt}
\setitemize[1]{itemsep=0pt,partopsep=0pt,parsep=\parskip,topsep=5pt}

\definecolor{darkgreen}{rgb}{0, 0.5, 0}

\let\oldcite\cite
\renewcommand{\cite}[1]{\textcolor{green!50!black}{\oldcite{#1}}}

\newcommand{\sys}{RASK\xspace}
\newcommand{\ebs}{EBS\xspace}
\newcommand{\vd}{VD\xspace}
\newcommand{\blockclient}{BlockClient\xspace}
\newcommand{\blockserver}{BlockServer\xspace}

\newcommand{\journal}{JournalFile\xspace}
\newcommand{\datafile}{DataFile\xspace}
\newcommand{\segmentcache}{SegmentCache\xspace}
\newcommand{\indexmap}{LBAIndex\xspace}
\newcommand{\offsettable}{CompressIndex\xspace}
\newcommand{\org}{Alibaba Cloud\xspace}

\newcommand{\rask}{RKey\xspace}
\newcommand{\pask}{point\xspace}
\newcommand{\plstree}{RASK\xspace}
\newcommand{\secondary}{SecondaryTree\xspace}
\newcommand{\unfound}{Unfound List\xspace}
\newcommand{\ltmap}{LT Map\xspace}
\newcommand{\nonoverlap}{NonOverlap List\xspace}
\newcommand{\representspace}{range space\xspace}
\newcommand{\sampled}{Sampled Dataset\xspace}
\newcommand{\full}{Full Dataset\xspace}
\newcommand{\origin}{Origin\xspace}
\newcommand{\compressunit}{CU\xspace}
\newcommand{\seqwritestream}{CW\xspace}
\newcommand{\cratio}{long CW ratio\xspace}
\newcommand{\codeword}[1]{\textcolor{black}{$\mathsf{#1}$}}
\newcommand{\myparagraph}[1]{\vspace{2pt}\noindent\textbf{#1}}

\definecolor{hightcode}{rgb}{0.6,0.1,0.1}
\definecolor{fcolor}{RGB}{93,49,49}
\definecolor{myRubineRed}{RGB}{209, 0, 86}

\newcommand{\highlightgreen}[1]{\color{ForestGreen}}

\usepackage[misc]{ifsym}
\usepackage{bbding}
\usepackage{breakurl}

\begin{document}

\pagestyle{plain}

\title{``Range as a Key'' is the Key! \\
Fast and Compact Cloud Block Store Index with {\sys}}

\author{Haoru Zhao$^{1}$,
Mingkai Dong$^{1}$, Erci Xu$^{1}$, Zhongyu Wang$^{2}$, Haibo Chen$^{1}$\\
{\normalsize {$^1$Shanghai Jiao Tong University}} \\
{\normalsize {$^2$Alibaba Group}} \\
}%

\sloppy

\maketitle

{\renewcommand{\thefootnote}{}\footnotetext{This is an extended version of the paper accepted at FAST '26.}}

\sloppy

\input{abs}

\setcounter{footnote}{0}

\input{intro}

\input{back}
\input{moti}

\input{trace}

\input{rkey}

\input{design}

\input{detail}

\input{impl}

\input{eval}

\input{related}

\input{concl}

\input{ack}

\appendix

\vspace{20pt}

\section*{Appendix}

\input{appendix/cu_exp}

\input{appendix/cu_alignment}

\input{appendix/split_proof}

\input{appendix/more_design}

\bibliographystyle{unsrt}
\bibliography{aliindex}

\end{document}

%% file: abs.tex
\begin{abstract}

In cloud block store, indexing is on the critical path of I/O operations and
typically resides in memory. 
With the scaling of users and the emergence of denser
storage media, the index has become a primary memory consumer, causing memory strain. 
Our extensive analysis of production traces reveals that write requests exhibit a strong tendency to target continuous block ranges in cloud storage systems. 
Thus, compared to current per-block indexing, our insight is that \emph{we should directly index block ranges (i.e., range-as-a-key) to save memory}.

In this paper, we propose {\sys}, a memory-efficient and high-performance
tree-structured index that natively indexes ranges. 
While range-as-a-key offers the potential to save memory and improve performance,
realizing this idea is challenging due to the range overlap and range fragmentation issues.
To handle range overlap efficiently,
{\sys} introduces the log-structured leaf, combined with range-tailored search and garbage collection.
To reduce range fragmentation,
{\sys} employs range-aware split and merge mechanisms.
Our evaluations on four production traces show
that {\sys} reduces memory footprint by up to 98.9\% and increases throughput by
up to 31.0$\times$ compared to ten state-of-the-art indexes.

\end{abstract}

%% file: intro.tex
\section{Introduction}%
\label{sec:intro}

Elastic Block Store (EBS) (e.g., Alibaba Cloud EBS~\cite{zhang2024ebs},
AWS EBS~\cite{AmazonEBS2023}, Azure Managed Disks~\cite{Azure2025,calder2011windows}, and Google Persistent Disk~\cite{google2023persistentdisk})
provides the virtual block device (VD) service and
serves as a cornerstone of 
modern cloud~\cite{satija2025cloudscape}.
Typically, EBS uses an index (EBS-index) to map logical block addresses (LBA) in 
VDs
to corresponding locations in the backend storage system (e.g., files in
the distributed file system, DFS). 
To ensure performance, the active EBS-index has to fully reside in memory.

However, 
as users scale up and larger storage media emerge (e.g., QLC SSDs~\cite{liang2019empirical}),
EBS-index now has to accommodate more entries
and becomes the dominant DRAM consumer, causing memory strains and further constraining physical storage utilization.
In
{\org}, a world-leading cloud vendor, its EBS-index consumes \textasciitilde57.1\% of the node's memory; 
in the most severe cases, \textasciitilde10\% of clusters
risk wasting \textasciitilde35\% of storage resources 
due to insufficient memory.

Given that the highly optimized EBS-index already outperforms other SOTA indexes in memory efficiency and performance (\autoref{fig:indexmap-comparsion}),
we start by revisiting the workload
characteristics in the field. 
We extensively analyze the large-scale EBS traces
provided by {\org}, 
spanning four clusters and eight representative application categories over one week.
One notable pattern is that individual writes that are close in time tend to have consecutive LBAs.
For example, given three write requests with LBAs [1,\,3]\footnote{In this paper, for a range [\emph{l}, \emph{r}], \emph{l} is the range's left bound (inclusive), \emph{r} is the right bound (inclusive).}, [9,\,10], and [4,\,7], the first and third are consecutive and can be merged into [1,\,7].

Such consecutive write sequences are widespread;
specifically, across eight reprsentative workloads in {\org} EBS traces, 65.0--81.5\% of writes are part of such sequences (\autoref{fig:sws_ratio}). We term such a sequence as a \emph{consecutive write} ({\seqwritestream}), which typically covers a block range (e.g., [1,\,7]).

Furthermore, writing to a range of blocks is not unique to {\org} or cloud block
store (i.e., {\ebs}).
We also observe frequent {\seqwritestream} occurrences in Tencent's {\ebs}
traces~\cite{zhang2020osca}.  
Cloud storage traces from Google~\cite{phothilimthana2024thesios} and Meta~\cite{pan2021tectonic,wang2024baleen} further reveal that writing to a
range is widespread in cloud storage systems.\footnote{Details for these four traces are provided in \autoref{sec:eval-setup}.} 
For example, 90.3\% of
writes span ranges exceeding four blocks in Meta's traces.  
We discover the \emph{root
cause} of this phenomenon is that both upper-level applications~\cite{vo2012logbase,mysql_log_2023,abadi2016tensorflow,shvachko2010hadoop,kreps2011kafka}\,(e.g., databases) and storage systems~\cite{rosenblum1992log,lee2015f2fs,linux_kernel_journaling,LinuxKernelPageCache,thekkath1997frangipani,anderson2020assise,yang2015split,whitaker2023scheduler}\,(e.g., filesystems and caching) tend to issue sequential writes. 

We validate this through white-box analysis using blktrace~\cite{blktrace} (\autoref{sub:call-for-range-as-a-key}).
Results show that the ratio of range writes is high (29.0--99.0\%), primarily caused by filesystem (FS) journaling and app-level services (e.g., Redis server, MySQL logs, etc.).

Given the prevalence of range writes, we argue that \emph{we should directly index ranges (i.e., \textbf{range as a key}) instead of individual blocks}.
For example, for the block range [1,\,7], we should index it with one entry instead of seven entries for each block.
Adapting range-as-a-key can reduce memory footprint by decreasing the number of entries, theoretically by 58.4\% to 91.1\% for EBS (\autoref{fig:io_compaction}).
It can also improve performance by (1) eliminating multiple index updates for a range and (2) speeding up queries due to the smaller index scale.

Thus, we propose \textbf{{\plstree}},
a high-performance and memory-efficient \textbf{\underline{R}}ange-\textbf{\underline{AS}}-a-\textbf{\underline{K}}ey tree index.
It consists of trie-format internal nodes and B-tree style leaves, where leaf entries represent ranges rather than individual objects.
{\sys} overcomes two key challenges to realize range-as-a-key efficiently.

\textbf{Challenge-1: Range overlap. }
Range overlap occurs when a new range (e.g, [2,\,4]) intersects with existing ones (e.g., [3,\,5]).
Range overlap degrades read performance as reads must identify the latest data from overlapping ranges.
Additionally, range overlap leads to memory waste 
by making old ranges obsolete when they are fully covered by newer ones
(e.g., [2,\,4] is covered by [1,\,5]);
however, removing these covered ranges upon their appearance requires complex operations and increases write latency.
Notably, the issues caused by range overlap cannot be resolved by adopting existing indexes, whether point-based (e.g., B-tree) or range-aware (e.g., interval tree~\cite{bentlay1979algorithms}, HINT~\cite{chris2022hint}).

To handle range overlap efficiently, we propose three techniques. 
First, we employ \emph{log-structured leaf} with append-only updates.
It batches the removal of covered ranges by garbage collection\,(GC) only when it is full, 
reducing the impact of overlap handling on write performance. 
Using the leaf as the GC unit also
enforces timely GC and
limits memory overhead.
Second, we design \emph{two-stage GC} to reduce write blocking time caused by GC.
The first stage quickly removes some common covered ranges to promptly resume writes, while the second stage thoroughly removes all fully covered ranges.
Third, we propose \emph{ablation-based search} to speed up lookups with overlapping ranges.
It scans the log-structured leaf in reverse order and ablates the target range gradually (i.e., focusing only on the unfound portions).
This efficiently avoids adding outdated values in the overlapping ranges to the result.

\textbf{Challenge-2: Range fragmentation. }
Since a leaf cannot contain an infinite number of entries to represent an infinite range space (i.e., key space),
range fragmentation will inevitably occur: User-written ranges have to be divided and stored in multiple leaf nodes if they span the range spaces of these leaf nodes.
This issue increases query, range management, and memory overhead.

We propose two techniques to mitigate range fragmentation.
We design \emph{range-conscious split} to reduce range fragmentation caused by leaf splits.
It tries to choose a split point that minimizes the number of ranges spanning the two new leaves, while balancing the entry count in these leaves.
Range fragmentation can also occur when newly inserted ranges do not align with the leaf's range space.
Therefore,
we introduce \emph{workload-aware merge and resplit} to dynamically adjust the leaf's range space
to better fit the workload,
further reducing fragmentation caused by new ranges.

\sloppy

We evaluate {\sys} against ten SOTA indexes with production traces from {\org}, Google, Meta, and
Tencent.  
Results show that {\sys} reduces memory footprint by 45.3--98.9\%, increases
throughput to 1.37--32.0$\times$, and reduces tail latency by 48.2--97.4\% on
average compared to baselines.  We further integrate {\sys} into
RocksDB~\cite{dong2021rocksdb} and evaluate its performance gains in KV store
scenarios (e.g., DFS metadata service).  RocksDB with {\sys} achieves up to
6.46$\times$ higher throughput than the original skiplist-based RocksDB.

In summary, we make the following contributions:
\begin{itemize}[leftmargin=1em,topsep=-0.14pt]
  \item \textbf{Analyses.} 
  Based on cloud storage trace analysis (e.g., {\org}), we find writes typically target a block range.
  \item \textbf{Indexing by range.} We propose that the index in cloud storage systems should
  use the block range as the key. 
  \item \textbf{{\sys}.} We design {\sys}, a memory-efficient and high-performance
   index that natively supports range-as-a-key. 
  Evaluations with traces
  from {\org} and others show its performance gains
  over SOTA indexes. 
\end{itemize}
We have open-sourced the code of {\sys} at \url{https://ipads.se.sjtu.edu.cn:1312/opensource/rask-index} and {\org} EBS traces at \url{https://tianchi.aliyun.com/dataset/218875}.

%% file: moti.tex
\section{Index Strains Memory in Modern \ebs}
\label{sec:moti}

\subsection{Background: Elastic Block Store}
\label{sub:elastic-block-storage}

\begin{figure}[t]
  \centering
  \includegraphics[width=\linewidth]{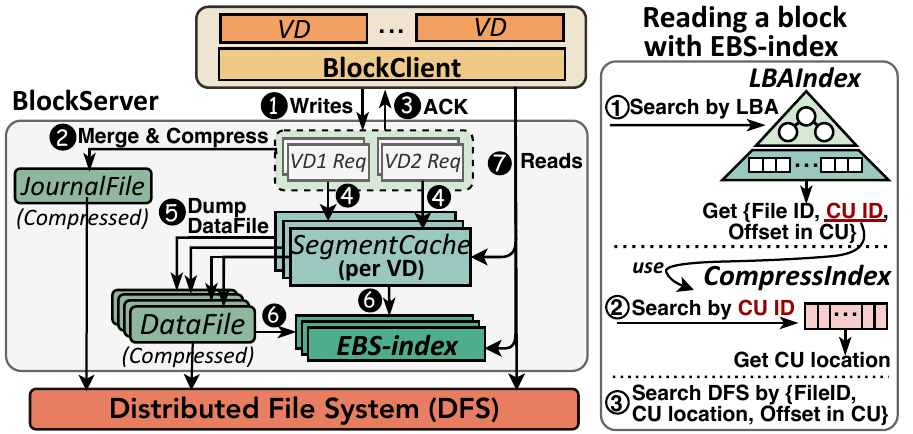}
  \vspace{-17pt}
  \caption{\textbf{{\ebs} architecture and read procedure.}}%
  \label{fig:ebs-arch}
  \vspace{-10pt}
\end{figure}

\myparagraph{Architecture.} 
Elastic Block Store ({\ebs}) is a core component of
cloud service, providing {\vd}s for
compute instances~\cite{AlibabaCloud2023,AmazonEBS2023,Azure2025,google2023persistentdisk}.
\autoref{fig:ebs-arch} shows the {\ebs} architecture
at Alibaba Cloud~\cite{zhang2024ebs}.
It consists of three layers: {\blockclient} in the compute layer, {\blockserver} in the proxy layer handling read/write requests from the {\blockclient}, and the DFS in the persistence layer.

\myparagraph{Write/Read.} 
For writes (\ding{182}), the path bifurcates. 
{\blockserver} first persists data in a {\journal} (\ding{183}) and then acknowledges the I/O completion (\ding{184}).
The uncompressed data is cached in the {\segmentcache} (\ding{185}).
Upon reaching a threshold (e.g., 512\,KiB), the cached data is written to
{\datafile}s in the DFS with compression (\ding{186}). 
To ensure the compression ratio and decompression efficiency, data is compressed in groups of four blocks, i.e., the compression unit ({\compressunit}) is four blocks.
The EBS-index is updated to map the VD's LBA to the
{\datafile} and the location in {\datafile} (\ding{187}).  
For reads, {\blockserver} first checks the {\segmentcache};
on a miss, it queries the EBS-index and then fetches data from DFS (\ding{188}). 

\myparagraph{Indexes in {\ebs}.} 
EBS-index includes in-memory {\indexmap} and {\offsettable}.
{\indexmap} maps LBA to the {\datafile} ID, {\compressunit} ID, and offset in the {\compressunit}.
It is a LSM-tree-like structure:
The top layer (i.e., MemTable) is in page-table format for fast LBA updates.
Once the MemTable's size reaches a threshold, it is converted into a 
more
memory-friendly immutable sorted array (i.e., SSTable), where each entry 
represents a write request's LBA range.
{\offsettable} maps {\compressunit} ID to {\compressunit}'s location in the {\datafile} using an array. 
When reading a block, as shown in \autoref{fig:ebs-arch},
{\ebs} first searches {\indexmap} by the target LBA (\ding{172}), then
uses the resulting {\compressunit} ID to
locate the {\compressunit} in {\offsettable} (\ding{173}), and finally reads DFS accordingly (\ding{174}).

\subsection{Memory Strain: Index is the Culprit}
{\org} Sysadmins report that EBS has been experiencing escalating
memory strains.
In the most severe cases, 
\textasciitilde10\% of clusters are at risk of wasting \textasciitilde35\% of physical storage resources.
The waste arises because these data cannot be indexed and thus become unusable under memory constraints.
This memory pressure is further intensified as EBS accommodates more users and adopts larger SSDs with denser NANDs.

We find that \emph{the {\ebs}-index is the main memory consumer,} which consumes thousands of TBs of memory for indexing thousands of PBs of data.  
{\org}'s statistics show that {\indexmap}
consumes \textasciitilde17.2\% memory, while {\offsettable} accounts for \textasciitilde39.9\%.
Indexes strain memory since: (1)
{\indexmap} scales with write requests, 
as a write request requires at least one SSTable entry. 
Besides, its log-structured format increases
memory overhead due to multiple versions.  
(2) {\offsettable}'s size is
proportional to the data volume due to per 4-block compression.

\begin{figure}[t]
  \centering
  \subfloat[Average memory usage per {\vd}]{
      \includegraphics[width=0.48\linewidth]{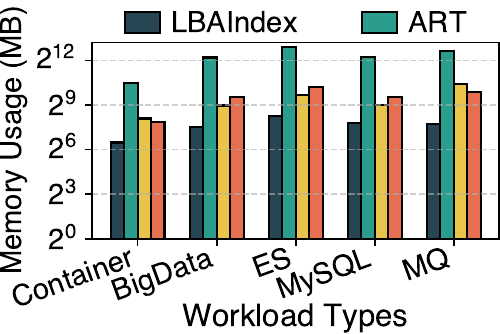}
      \vspace{-6pt}
      \label{fig:indexmap-memory}}
  \subfloat[Average throughput per {\vd}]{
          \includegraphics[width=0.48\linewidth]{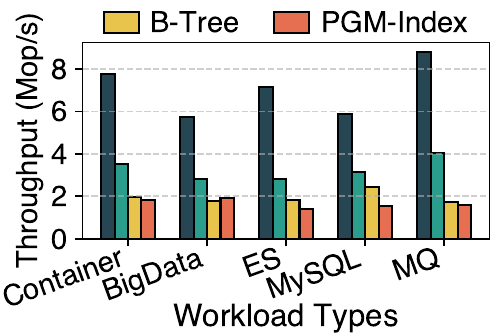}
          \vspace{-6pt}
          \label{fig:indexmap-throughput}}
  \caption{\textbf{Comparison of {\indexmap} and SOTA indexes.} In
  this paper, \emph{ES} is Elasticsearch, \emph{MQ} is message queue.}%
  \label{fig:indexmap-comparsion}
\end{figure}

\subsection{Exploration of Possible Remedies}
\label{sub:explore-solution}

\myparagraph{Modifying the architecture.}
One may ask if memory strain can be reduced by modifying EBS architecture (e.g., removing {\segmentcache}, writing directly to {\datafile}) or simply adding more memory. 
However, the memory overhead of EBS-index stems fundamentally from the indexed data volume, which architectural changes cannot resolve.
Adding memory is costly, inflexible, and unsustainable in the non-stop production environment, as it needs a time-consuming and laborious maintenance beyond just the memory cost.

\myparagraph{Replacing index structure.} 
Another possibility is to use a more
memory-efficient index that supports in-place updates, thereby avoiding the overhead from
multiple versions.    
To evaluate this idea, we compare {\indexmap} with B-tree~\cite{bingmann_stx_btree},
SOTA memory-efficient trie (adaptive radix tree, ART~\cite{leis2013adaptive}), and learned index
(PGM-Index~\cite{ferragina2020pgm}) under five typical {\ebs} workloads.
\autoref{fig:indexmap-comparsion} shows that {\indexmap} already consumes less
memory and performs better than these SOTA indexes.  
This is because: 
(1) {\indexmap}'s MemTable is size-limited and SSTables index at
write-request granularity, while other indexes use LBA granularity; 
(2) MemTable updates have O(1) time complexity thanks to its page-table format.  

\myparagraph{Applying larger compression unit (CU).} 
{\ebs} uses a short CU (i.e., 4
blocks) to avoid severe read amplification during decompression. This is because
{\ebs} adopts a stream-based compression scheme~\cite{google_snappy,lz42025} (e.g., LZ4) which requires
sequential decompression from the CU start. 
A possible alternative is to
use a random-access compression scheme, which allows larger CUs by avoiding full decompression from the start.
However, such schemes are inefficient and fail to decrease the volume of information to be indexed:
(1) To support random access, these schemes need to build a
dictionary~\cite{zraorg2023,robert2006new}, unfriendly to the latency-sensitive
EBS. 
For instance, we explore with a SOTA method called
FSST~\cite{boncz2020fsst}, which takes \textasciitilde1\,ms to build the dictionary and compresses at a rate 9.78$\times$ slower than LZ4.  
(2) Compressed blocks have variable lengths; therefore,
we still need to record their offset information within the compressed data for
decompression, eliminating the memory benefits of introducing larger CUs.

%% file: trace.tex
\section{Characterizing Block Service Access Pattern}
\label{sec:trace}

\begin{figure}[t]
  \centering
  \begin{minipage}[b]{0.16\textwidth}
      \centering
      \includegraphics[width=\textwidth]{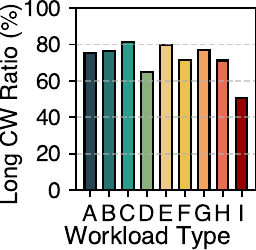}
      \vspace{-13pt}
      \subcaption{Long {\seqwritestream} ratio}
      \label{fig:sws_ratio}
  \end{minipage}
  \hfill
  \begin{minipage}[b]{0.29\textwidth}
    \centering
    \includegraphics[width=\textwidth]{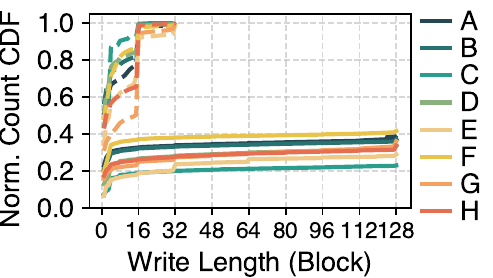}
    \vspace{-15pt}
    \subcaption{{CDF of origin/compacted write length}}
    \label{fig:io_compaction}
\end{minipage}
  \vspace{-7pt}
  \caption{\textbf{Write pattern analysis.} 
  (b) Write length CDF before (dashed lines) vs. after (solid lines) I/O compaction; y-axis is normalized to pre-compaction write count. 
  Workloads are: \emph{Container}\,(A), \emph{Message Queue}\,(B), \emph{MySQL}\,(C), \emph{Elasticsearch}\,(D), \emph{Redis}\,(E), \emph{MongoDB}\,(F),
  \emph{BigData}\,(G), \emph{WebApp}\,(H), \emph{External}\,(I). 
  }
  \label{fig:seq_write_total}
  \vspace{-5pt}
\end{figure}

The above analysis indicates that the memory pressure in {\ebs} cannot be resolved by altering the architecture, modifying the index structure, or adopting alternative compression schemes.
Hence,
we analyze the field traces in the hope of exploiting the workload
characteristics to reduce the memory footprint. 

We get two datasets recently collected by {\org} EBS:
(1) one-week I/O traces of 1.4\,k VDs in two clusters; 
(2) three-day I/O traces of 400 VDs in two other clusters. 
They cover representative applications like
databases, KV stores, message queues, big data processing, search
engines, containers, and web apps.

\subsection{Write-Write Correlation}
\label{sub:write-write-correlation}

\noindent
\textbf{Observation.} 
\emph{Individual write requests that are temporally close can be
spatially consecutive.} 
E.g., 
four write requests (A, B, C, D) target LBAs [1,\,4],\,[11,\,12],\,[5,\,6], and\,[7,\,8], respectively.
While A, C, and D are independent requests, their LBAs are consecutive (i.e., [1,\,4], [5,\,6], and [7,\,8] can be concatenated).
To quantify this pattern, we define an \emph{observation window} as the scope for checking consecutiveness.
If the window covers 4 writes (i.e., we check consecutiveness among every 4 write requests), then in this example, A, C, and D form a consecutive sequence.
We define such a consecutive write sequence within the observation window as a \emph{Consecutive Write} ({\seqwritestream}).
For a {\seqwritestream} with $\geq$ 2 requests, we classify it as a long {\seqwritestream}.
We further define \emph{\cratio} as the proportion of write requests that belong to long {\seqwritestream}s.
The {\cratio} of this example is 75\% (i.e., A,\,C, and\,D out of 4 requests).

As shown in \autoref{fig:sws_ratio}, the {\cratio}s are consistently high (65.0--81.5\%) across all workloads when the window is 36 write requests.
\footnote{Similar phenomena are observed across windows of 2-128 requests, with detailed figures provided in the \autoref{sec:cu-exp} for reference.}
This phenomenon also shows in Tencent's {\ebs} traces~\cite{zhang2020osca}
(\emph{External}\,(I) in \autoref{fig:sws_ratio}). 
The prevalence of long {\seqwritestream}s renders a memory-saving opportunity by consolidating multiple entries into one for a {\seqwritestream} in the {\indexmap}.

\myparagraph{Optimization: I/O compaction.}
The first optimization is to enlarge the {\indexmap}'s granularity to {\seqwritestream}s.
We can leverage the {\segmentcache} (\autoref{fig:ebs-arch}) to capture {\seqwritestream}s,
using its size\footnote{Given the average write length is 3.63 blocks at  {\org}'s {\ebs}, this window size is roughly equivalent to 35.2 write requests.} (128 blocks) as the observation window.
When flushing the {\segmentcache}, we reorder/merge these cached writes into {\seqwritestream}s and then write these {\seqwritestream}s to DFS.
This upsizes the granularity of {\indexmap} from individual requests to {\seqwritestream}s, reducing the entry count and lowering the memory footprint.

\myparagraph{Potential benefit.}
\autoref{fig:io_compaction} compares the cumulative distribution function (CDF) of write lengths with/without I/O compaction.
Results show that I/O compaction reduces {\indexmap} write counts by 58.4--77.0\%, greatly reducing the number of {\indexmap} entries.
Moreover, the reordering/merging overhead is negligible since:
(1) The {\segmentcache} is small;
(2) It only manipulates metadata (i.e., LBAs and block positions in cache) without interfering with DFS write processes.

\begin{figure}[t]
  \centering
  \includegraphics[width=0.99\linewidth]{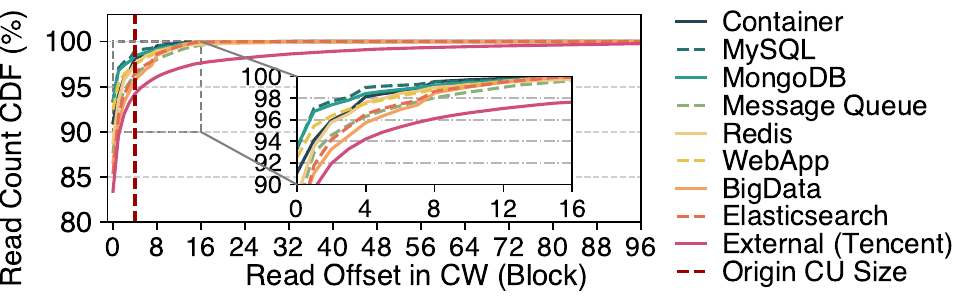}
  \vspace{-5pt}
  \caption{\textbf{CDF of read start offset relative to
  the associated {\seqwritestream}.} }
  \label{fig:read_alignment}
  \vspace{-5pt}
\end{figure}

\subsection{Write-Read Correlation}
\label{sub:write-read-correlation}

\myparagraph{Observation.}
\emph{When reading data written by a {\seqwritestream}, the request nearly always starts from the {\seqwritestream} beginning.}
E.g., if a {\seqwritestream} covers LBAs [1,\,8],
the following reads tend to start from its beginning (i.e., 1), while 
the likelihood of starting from the later part (i.e., 5--8) is low.  
\autoref{fig:read_alignment} shows that across 8 workloads, 
> 85.4\% of reads align with the {\seqwritestream} start, 

while <\,1\% of reads begin 10 blocks away from the {\seqwritestream} start.
Tencent's traces also show a similar pattern, i.e., 83.4\% of reads align with the {\seqwritestream} start.
Intuitively, this pattern reflects how an app writes implies how it would read,
e.g., 
a {\seqwritestream} for logging is likely to be read from the start.
Notably, over 95.7\% of reads start within 4 blocks (i.e., the {\org} EBS {\compressunit} size) of the {\seqwritestream}'s beginning.
It means that enlarging CU to {\seqwritestream} causes negligible read amplification during decompression. 

\myparagraph{Optimization: CU alignment.}
The second optimization is to align the CU with the {\seqwritestream}. 
We propose an adaptive compression scheme: {\seqwritestream}s exceeding four blocks are compressed
per {\seqwritestream}, while shorter {\seqwritestream}s use the 4-block CU to
maintain the compression ratio.
It expands the CU indexing granularity from 4-block to {\seqwritestream}s, reducing the memory footprint. 

\myparagraph{Potential benefit.}
After applying CU alignment, the number of CUs to be indexed reduces by 69.1--91.1\% across eight workloads.
Additionally, the overhead of CU alignment is acceptable.
\autoref{fig:read_alignment} indicates that
\textasciitilde95.7\% requests are basically unaffected, while the increased
latency for the remaining ones is in an acceptable range (0.477\%--2.60\%).
\footnote{Details of the CU alignment implementation and its evaluation workflow are provided in supplementary material S2 for reference.}

\begin{figure}
  \centering
  \includegraphics[width=0.75\linewidth]{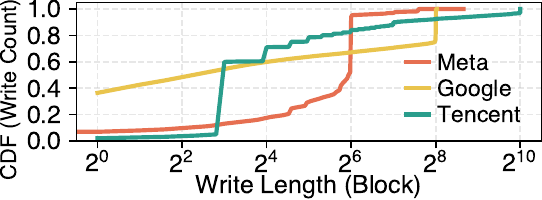}
  \vspace{-5pt}
  \caption{\textbf{Write length distribution of industry
  traces.}}
  \label{fig:industry_write_length}
  \vspace{-5pt}
\end{figure}

\subsection{Call for ``Range as a Key''}
\label{sub:call-for-range-as-a-key}

The memory savings from I/O compaction and CU alignment rely on enlarging index granularity to {\seqwritestream}s.
In this case, the original EBS-index is not suitable because:
(1) As shown in \autoref{fig:io_compaction}, {\seqwritestream}s are often long LBA ranges but the {\indexmap}'s MemTable uses single-LBA as index granularity.
(2) {\offsettable} only supports fixed-length {\compressunit}s and not applicable for variable-length {\seqwritestream}s.
Thus,
an index that uses \textbf{\emph{range as a key} ({\rask})} is essential to achieve memory savings.
This index uses the LBA range of a {\seqwritestream} as the key,
mapping it to the corresponding DFS location and compression metadata.

Furthermore, the benefits of range-as-a-key are beyond just {\org} and/or {\ebs}.
Based on the following analysis, we find that \emph{using range-as-a-key is advantageous for various cloud storage systems with range-write heavy workloads.}

%% file: rkey.tex
\input{rkey/general}

\input{rkey/alternative}

%% file: rkey/general.tex
\myparagraph{Common root causes.}
{\seqwritestream} exists primarily because: (1) Both FSs and applications prefer sequential
writes to match block device characteristics (e.g., sequential writes on
HDD/SSDs are faster than random writes);
(2) 
Multi-app interactions in the system interrupt an app's sequential writes, creating multiple
{\seqwritestream}s.
We confirm this via white-box analysis
using blktrace~\cite{blktrace}. 
Specifically, we analyze 1-hour I/O samples of MySQL and Redis running TPC-C and YCSB workloads, respectively.
Results show that 
long {\seqwritestream} ratios
are high (29.0--99.0\%), primarily caused by
 FS journaling and application
services (e.g., Redis server, MySQL InnoDB log, etc.).

\myparagraph{Prevalence in cloud storage.}
Previously, we have shown similar patterns in Tencent traces.  
Further, by
analyzing traces from Google~\cite{phothilimthana2024thesios},
Meta~\cite{pan2021tectonic,wang2024baleen}, we
find that in their cloud storage systems, while block is the basic operation unit,
\emph{writes typically span a range of blocks.}
As \autoref{fig:industry_write_length} shows, 51.6\% (Google) and 90.3\% (Meta) of writes exceed four blocks.
Thus, {\rask} is not limited to a 
specific system or company.

Given the prevalent range writes, indexing by block ranges (start LBA + write length)
instead of individual blocks offers:
(1) \emph{Memory efficiency}: reduces N\,-\,1 entries per N-block range.
(2) \emph{Performance improvement}: avoids multiple index updates for a range
and speeds up queries via a smaller index scale.

\myparagraph{Possible extensions of {\rask}.}
The wide existence of range writes implies the deployment of {\rask} is likely to benefit various scenarios. 
(1) \emph{Example-1:\,Flash cache.} In bulk
storage systems, flash cache stores hot data in SSDs to reduce HDD
load~\cite{yang2022cachesack,mcallister2024fairywren}.  
It needs a DRAM
index to map block locations in SSDs.  
Compared to existing B-tree
indexes~\cite{berg2020cachelib}, {\rask} can reduce index's memory footprint, which
is a key concern in flash cache~\cite{mcallister2021kangaroo}.
(2) \emph{Example-2:\,DFS metadata service.} Many DFSs use KV stores to manage
metadata (e.g., LSM-tree~\cite{pan2021tectonic,ren2014indexfs},
B-tree~\cite{li2017locofs}).  
To speed up metadata operations, it stores block-to-file mappings (e.g., Tectonic,
Meta's exabyte-scale DFS).
Here,
{\rask} is more efficient for block indexing as it avoids one-by-one
updates and reduces entry count, allowing more entries to reside in memory for
faster queries.

%% file: rkey/alternative.tex
\subsection{Range-as-a-key: Not off-the-shelf}
\label{sub:not_off_the_shelf}

\begin{figure}[t]
  \centering
  \includegraphics[width=0.95\linewidth]{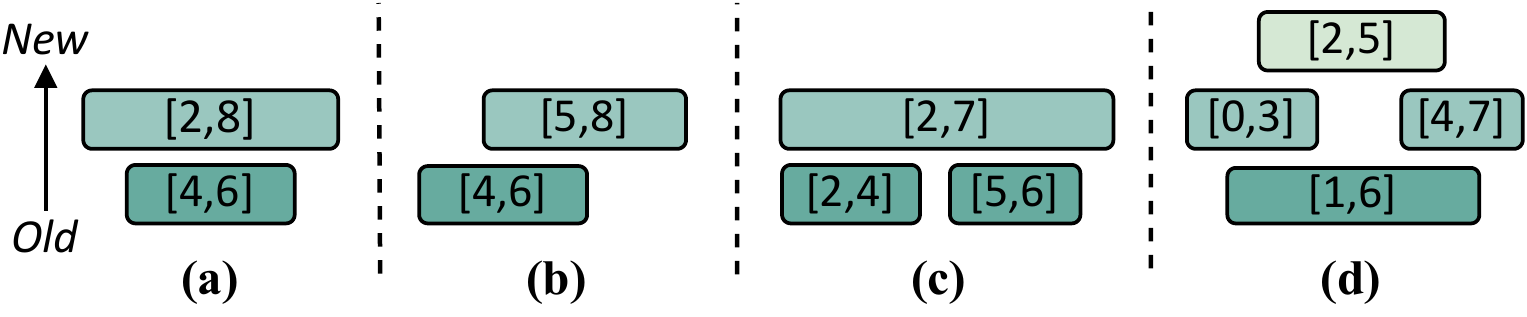}
  \vspace{-10pt}
  \caption{\textbf{Range overlap cases.} 
  }%
  \label{fig:range-overlap}
  \vspace{-5pt}
\end{figure}

Compared to point indexes (e.g., B-tree), indexing ranges is more challenging since it must
handle various range overlap cases.  
\autoref{fig:range-overlap} shows some range overlap cases: (1)\,cover an old range\,(\autoref{fig:range-overlap}(a)); (2) partially overlap an old range\,(\autoref{fig:range-overlap}(b)); (3)\,cover/partially overlap many old ranges\,(\autoref{fig:range-overlap}(c,\,d)).
An efficient {\rask} index should promptly remove fully covered ranges to avoid memory waste and quickly locate the latest data in the overlapping ranges during reads. 
However, neither existing range-aware indexes nor adapted point indexes can meet these requirements well, as we will show next.

\begin{figure}[t]
  \centering
  \subfloat[Memory usage]{
      \includegraphics[width=0.3\linewidth]{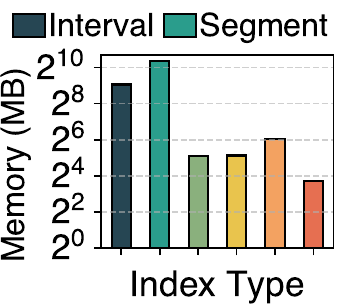}
      \vspace{0pt}
      \label{fig:range_tree_memory}} \subfloat[Throughput]{
      \includegraphics[width=0.667\linewidth]{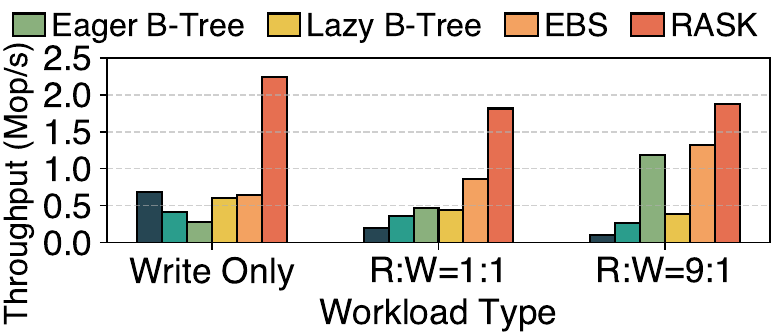}
      \vspace{-5pt}
      \label{fig:range_tree_tput}}
   \caption{\textbf{Comparison of existing range-aware indexes, EBS-index, and
   {\sys} (our design).} \emph{Eager} and \emph{Lazy B-tree} are adapted using the eager and lazy methods. EBS is EBS-index. 
   The dataset is 10M ranges with uniformly distributed lengths 4--64.
   }
  \label{fig:range_tree}
\end{figure}

\myparagraph{Porting the range-aware indexes?}
Indexing ranges is common in temporal databases,
with typical indexes like interval tree~\cite{edelsbrunner1980},
1D-grids~\cite{beckmann1990rtree},  segment
tree~\cite{deBerg2008}, HINT~\cite{chris2022hint}, etc.~\cite{periodindex2019,kaufmann2013timeline,kriegel2000managing,bouros2017forward,bouros2021memory,djingos2014overlap,piatov2021cache,Chaabouni1993ThePT,Hanson1991TheIS,huiba2020dadi}.  
However, \emph{these
indexes mainly focus on intersection queries, without removing covered ranges automatically.}
This is because they primarily target secondary index scenarios,
where the covered ranges are still useful and cannot be deleted
(e.g., record
[$k_1$,\,$v_1$] with time range [$t_1$,\,$t_2$] is not overwritten by [$k_2$,\,$v_2$] with [$t_1$,\,$t_3$]).  
But in
our case, 
not removing covered ranges wastes
memory and degrades performance.  
Thus, as shown in \autoref{fig:range_tree}, interval tree and segment tree are
inefficient in both memory and read performance. 

\myparagraph{Adapting the point indexes?}
When adapting point indexes to {\rask} by indexing the range's left bound, there are two possible ways.
(1) \emph{Eager}: Writes find all
overlapping ranges and remove their intersecting portions before insertion; 
reads find all intersecting ranges with the target range. 
(2) \emph{Lazy}: Writes insert ranges with sequence numbers and only remove
covered ranges; 
reads find all intersecting ranges and select the
ones with the highest sequence number for each overlapping position.  
However, \emph{both
methods perform poorly due to the high cost of handling range overlaps.}
\autoref{fig:range_tree_tput} shows that Eager and Lazy B-tree perform much
worse than EBS-index under write-heavy and read-heavy workloads, respectively.

%% file: design.tex
\section{{\sys} Overview}
\label{sec:challenge_and_overview}

\begin{figure}[t]
  \centering
  \includegraphics[width=0.94\linewidth]{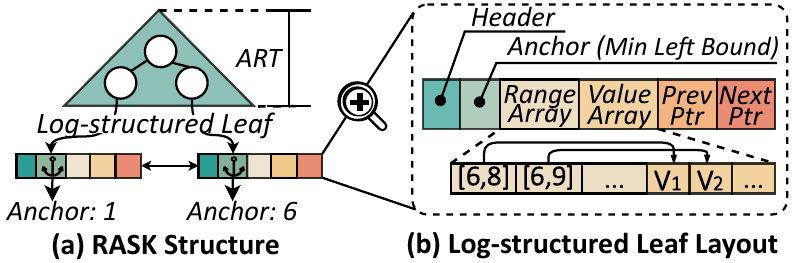}
  \caption{\textbf{The structure of {\sys}.}}%
  \label{fig:pls_arch}
\end{figure}
We propose {\sys}, a high-performance and memory-efficient in-memory index that natively supports range-as-a-key.
Based on the analysis in \autoref{sub:call-for-range-as-a-key}, \emph{{\plstree} is a general-purpose index for range-write intensive workloads, not just for the {\ebs} at {\org}.} 
It offers a general range read/write interface (\autoref{sub:op}), potentially benefiting more systems (e.g., flash cache, DFS metadata service).

\subsection{Structure and Workflow}
\label{sub:arch}

\myparagraph{{\plstree} structure.}
As \autoref{fig:pls_arch}(a) shows,
{\plstree} uses ART~\cite{leis2013adaptive} for internal nodes and employs the \emph{log-structured leaf} (i.e., leaves are globally ordered, but updates within the leaf are append-only).
ART is a trie variant that indexes keys by storing them as paths of characters, with each internal node representing a common key prefix.
We choose ART since it is more efficient than B-tree, while also being memory-friendly via path compression and internal node resizing.
The log-structured leaf allows for efficient range overlap handling.

\autoref{fig:pls_arch}(b) shows the leaf layout.
Each leaf is identified by an \emph{anchor key} that represents the minimum left bound of its ranges, and this anchor key is indexed by internal nodes.
A leaf's {\representspace} covers all ranges\footnote{In this paper, \emph{range key} is defined as \emph{range left bound}, and \emph{range length} = \emph{range right bound (inclusive)} - \emph{range left bound} + 1.} whose left bound $\ge$ its anchor key and right bound $<$ the next leaf's anchor key, which is non-overlapping with other leaves.
For each entry, the range (i.e., key) is stored in the \emph{Range Array} and the value is stored in the \emph{Value Array}.
Leaves are doubly linked to support efficient range operations.
Additionally, each leaf contains an 8-byte header to store current entry count and concurrency control information (\autoref{sub:concurrency}).

\myparagraph{Basic workflow.}
For reads, 
{\plstree} first traverses internal nodes to locate the target leaf (i.e., the last leaf whose anchor key $\leq$ the target range's left bound).
Then it extracts the latest value from the intersection of the target range and the leaf's ranges.
For example, to read the range [7,\,8] from \autoref{fig:pls_arch}(a), it first locates the leaf with anchor key\,=\,6, then retrieves value for [7,\,8] from entry [6,\,9] in that leaf (not from [6,\,8]).
For writes, {\plstree} first locates the target leaf as reads, then appends new entries to this leaf's range and value arrays.
If this leaf is full (i.e., entry count reaches the capacity), GC is triggered to remove old ranges that are fully covered by newer ones.
E.g., for the leaf in \autoref{fig:pls_arch}(b), [6,\,8] can be reclaimed by GC.
If this leaf remains full after GC, it is split, and internal nodes are updated with the new leaf's anchor key.

\subsection{Challenges and Approaches}

{\sys} tackles two key challenges from {\rask}: range overlap and fragmentation, achieving efficient read/write and structural modification operations (SMO, e.g., split/merge).

\myparagraph{Challenge-1: Range overlap.}
When writing a new range, range overlaps may cause some old ranges to be fully covered (\autoref{fig:range-overlap} (a),\,(c),\,(d)).
If these ranges are not removed in time, they will waste memory and degrade read performance due to the need to check more ranges.
However, promptly removing covered old ranges upon their appearance degrades write efficiency due to the complexity of identifying covered ranges.
Additionally, overlapping ranges harm read performance, as reads must identify the latest data of the overlapped portions.

\myparagraph{Addressing Challenge-1 in write and read. }
{\plstree}'s \emph{log-structured leaf} underpins efficient range overlap handling,
benefiting both writes and reads.
For writes, the log-structured design 
reduces the impact of removing covered ranges by batching them during GC, which is triggered only when the leaf is full.
Moreover, using the leaf as a fine-grained GC unit ensures prompt GC, thereby limiting the memory overhead.
For reads, the log-structured layout enables early termination once the target range is fully retrieved.
To further optimize searches with overlapping ranges in a leaf, we propose \emph{ablation-based search} (\autoref{sub:search}).
For GC, we introduce \emph{two-stage GC} (\autoref{sub:gc}) to reduce the write blocking time caused by GC while removing fully covered ranges as much as possible.

\begin{figure}[t]
  \centering
  \includegraphics[width=0.99\linewidth]{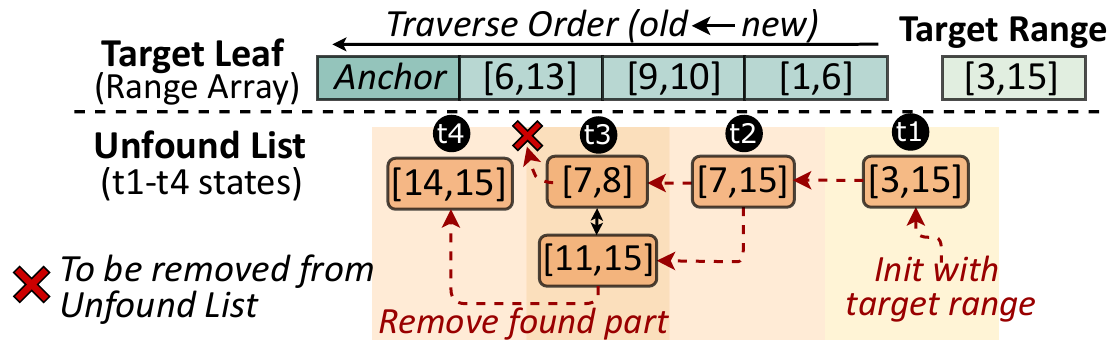}
  \vspace{-10pt}
  \caption{\textbf{Ablation-based search example.}}%
  \label{fig:search}
  \vspace{-10pt}
\end{figure}

\myparagraph{Challenge-2: Range fragmentation.\,}
Given that a leaf cannot have an infinite capacity to represent the infinite range space, range fragmentation is inevitable to occur:
A user-written range that spans multiple leaves' {\representspace} is divided and stored in these leaves.
This issue increases read/write, range management, and memory overhead.

\myparagraph{Addressing Challenge-2 in split and merge. }
{\sys} reduces range fragmentation through proper split and merge/resplit operations.
We employ \emph{range-conscious split} (\autoref{sub:split}) to mitigate 
dividing the leaf's ranges during leaf splits
while balancing the entry count in new leaves.
Range fragmentation can also occur when newly user-written ranges do not align with existing leaves' range space.
To resolve this issue, we introduce \emph{workload-aware merge and resplit} (\autoref{sub:merge}), which dynamically adjusts leaves' {\representspace} to adapt to workload characteristics, reducing fragmentation-induced overhead.

%% file: detail.tex
\section{{\sys} Design}
\label{sec:design}

We have achieved efficient range writes with the log-struc- tured leaf (\autoref{sub:arch}).
In this section, we present how {\sys} support efficient read (\autoref{sub:search}), GC (\autoref{sub:gc}), and SMOs (\autoref{sub:split}, \autoref{sub:merge}).

\input{design/search}

\input{design/gc}

\input{design/smo}

%% file: design/search.tex
\subsection{Ablation-based Search}
\label{sub:search}

{\plstree}'s search procedure traverses the leaf in reverse order until the target range is fully retrieved or all entries are processed.
For each range in a leaf, 
{\sys} collects its intersection with the unfound portions of target range.
For example, when searching for range [3,\,15] in \autoref{fig:search}'s leaf, the range [1,\,6] contributes [3,\,6] to the result.
The key challenge of search lies in efficiently maintaining the target range's search state.

Given that this search procedure can be seen as ablating the target range with the leaf's ranges, 
we propose an \emph{ablation-based search}. 
It tracks the search state using an ordered list of target range's unfound subranges (i.e., \emph{{\unfound}}).
Specifically, the {\unfound} is initialized with the target range (e.g., [3,\,15] at t1 in \autoref{fig:search}).
Then for each range $R$ in leaf, 
the intersections of $R$ and {\unfound} (i.e., 
$\{R \cap R'$ $\mid$ $R'$ $\in$ $\text{\unfound}\}$) are retrieved and removed from {\unfound} 
(e.g., at t2, [3,\,6] is removed from [3,\,15], leaving [7,\,15]). 
As the search proceeds, {\unfound} is gradually ablated.

In this scenario,
the ordered list is well-suited for tracking the unfound subranges.
\unfound's ordered property helps efficiently locate subranges that overlap with $R$ when getting the intersection of $R$ and {\unfound}.
Specifically,
once the first overlapping subrange is found (e.g., [7,\,8] for leaf range [6,\,13] at t3), all subsequent overlapping ranges (e.g., [11,\,15] at t3) can be accessed sequentially.
Then the intersections can be removed with O(1) time complexity.
While locating the first overlapping range in {\unfound} requires linear search, 
the cost of linear search is bounded, as the number of unfound subranges is limited by processed entry count of the leaf.

%% file: design/gc.tex
\begin{figure}[t]
  \centering
  \begin{subfigure}[b]{0.91\linewidth}
    \centering
    \includegraphics[width=\linewidth]{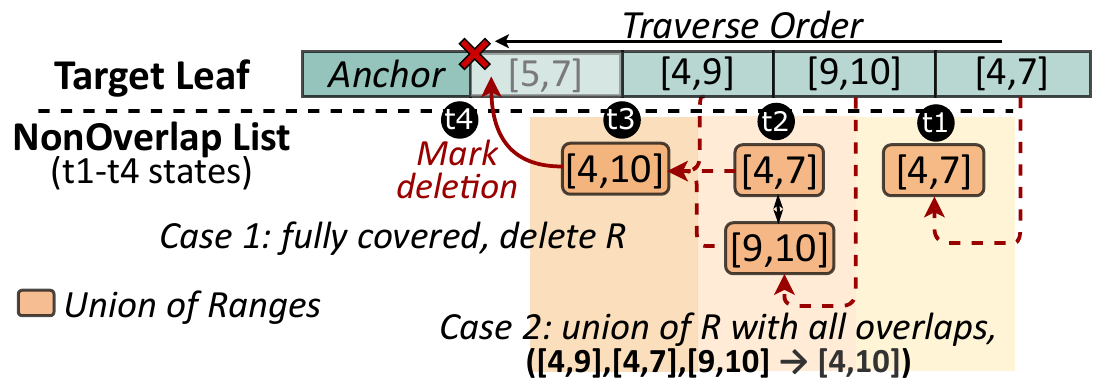}
    \vspace{-15pt}
    \caption{Normal GC example. R is range.}
    \label{fig:thorough_tidy}
  \end{subfigure}
  \vspace{5pt}
  \begin{subfigure}[b]{0.96\linewidth}
    \centering
    \includegraphics[width=\linewidth]{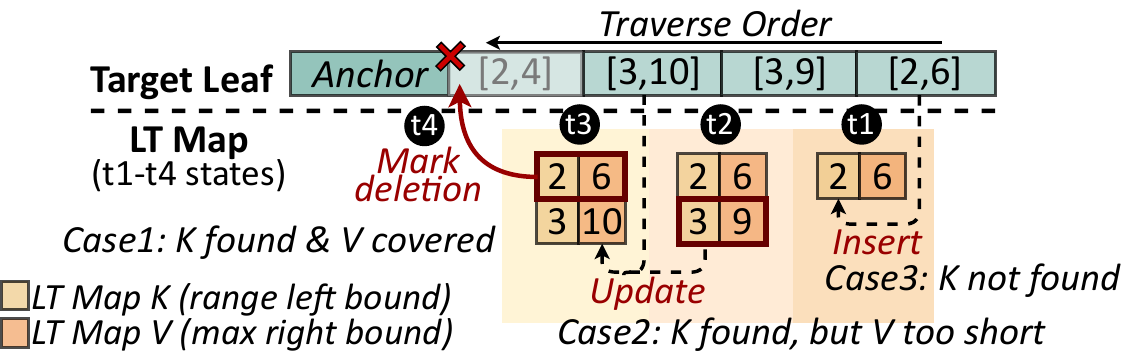}
    \vspace{-15pt}
    \caption{Lightweight GC example.}
    \label{fig:lightweight_tidy}
  \end{subfigure}
  \vspace{-14pt}
  \caption{\textbf{Two-stage GC examples.}}
  \label{fig:gc_examples}
  \vspace{-5pt}
\end{figure}

\subsection{Two-stage GC}
\label{sub:gc}
If a leaf is full, GC frees up space by removing fully covered old entries.
GC efficiency is crucial to write performance, as writes must wait for GC to complete.
The key challenge of GC is identifying old ranges covered by the union of multiple new ranges (e.g., [1,\,6] in \autoref{fig:range-overlap}(d) is covered by three ranges).

To address this, 
we perform GC via reverse-order leaf scanning while maintaining the union of all processed ranges to check if a preceding range is covered (termed as \emph{normal GC}).
Specifically, we employ an ordered list (i.e., \emph{\nonoverlap}) to track the union of processed ranges, stored as multiple non-overlapping ranges.
As shown in \autoref{fig:thorough_tidy}, for each range $R$, we check if $R$ is fully covered by an entry in {\nonoverlap}.
If so, $R$ is marked for deletion (case\,1\,at\,t4: [5,\,7] is covered by [4,\,10]).
Otherwise, {\nonoverlap} is updated by taking the union of $R$ with all overlapping ranges in {\nonoverlap} (case\,2\,at\,t3: [4,\,10] is the union of [4,\,9], [4,\,7] and [9,\,10]).

Normal GC is efficient in most cases, where the \nonoverlap only has a few entries.
However, when the number of entries in {\nonoverlap} is large, the linear search for identifying overlapping ranges can still be a bottleneck.
We seek to further improve GC efficiency by avoiding the linear search.
Fortunately, we find that many ranges are fully covered by new ranges with the same left bound.  
Specifically, we replay the EBS traces from {\org} with only normal GC enabled in {\sys}. 
Among all reclaimable entries, we find that on average 73.8\% of them can be reclaimed by only checking newer ranges with the same left bound.

Therefore, we further propose a \emph{lightweight GC} that only checks if ranges are fully covered by newer ones with the same left bound.
As \autoref{fig:lightweight_tidy} shows, \emph{lightweight GC} scans the leaf's range array backwards and maintains an O(1)-complexity map (i.e., \emph{{\ltmap}}) to track the maximum right bound per left bound.
For each range $R$, if its left bound exists in {\ltmap} and right bound $\leq$ the recorded value (case\,1 at t4), $R$ is marked for deletion; 
otherwise, {\ltmap} is updated with $R$'s left and right bounds (case\,2 at t3\,\&\,case\,3 at t1).
Lightweight GC avoids the linear search and can quickly reclaim some space, further reducing write blocking time.

In summary, we design a two-stage GC to reduce the write blocking time:
(1) \emph{Lightweight GC}: remove some fully covered entries quickly;
(2) \emph{Normal GC}: remove all fully covered entries and is triggered only when lightweight GC cannot free up any space.
After GC completes checking all entries, those marked for deletion are removed, and the remaining ones are shifted forward to fill holes and keep the original order.

%% file: design/smo.tex
\subsection{Range-conscious Split}
\label{sub:split}

\begin{figure}[t]
  \centering
  \includegraphics[width=0.49\textwidth]{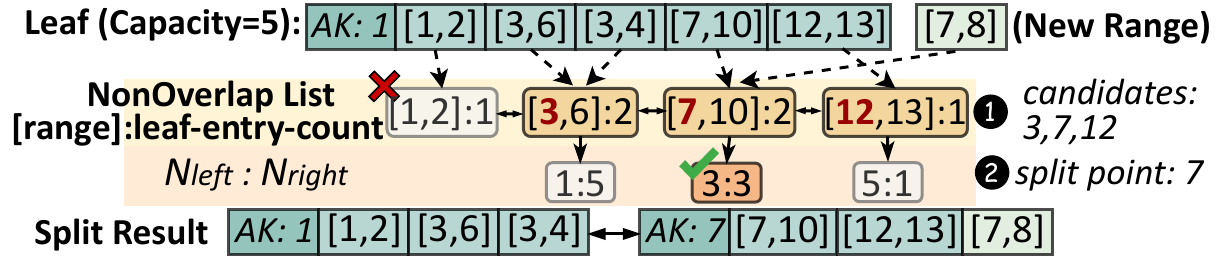}
  \vspace{-15pt}
  \caption{\textbf{Split example.} AK is anchor key.}%
  \label{fig:normal_split}
  \vspace{-5pt}
\end{figure}
When inserting into a full leaf and GC cannot free up space,
the leaf is split at the split point $P_{s}$ to create a new right leaf ($L_{r}$).
Entries with range left bounds $\geq$ $P_s$ are moved to $L_{r}$,
and entries across the $P_s$ 
are divided and stored in both leaves.
Then internal nodes are updated with $L_{r}$'s anchor key.
The key difference between {\plstree} and point indexes (e.g., B-tree) in split is that:
The $P_{s}$ should not only try to balance entry counts between two new leaves, but also avoid range fragmentation (i.e., $P_{s}$ should not intersect with any ranges in the leaf). 

To avoid range fragmentation, 
\ding{182} we use the left bounds of all entries in GC-obtained {\nonoverlap} (except the first one) as $P_{s}$ candidates (e.g., 3, 7,\,12 in \autoref{fig:normal_split}).
This is because these entries are non-overlapping, and their boundaries naturally guarantee no intersection with the leaf's ranges.
To achieve better balance,
\ding{183} we choose the $P_s$ candidate that most evenly balances entry count between two new leaves (e.g., 7 in \autoref{fig:normal_split} has three entries on the left and three on the right).
This can be calculated according to the entry count information collected during the {\nonoverlap} construction.

When {\nonoverlap} cannot provide any candidates (i.e., it has only one entry), 
we directly select $P_{s}$ from the boundaries of leaf's ranges for efficiency (e.g., for \autoref{fig:normal_split}'s leaf, the boundaries are 1,\,2,\,3,\,$\ldots$,\,10,\,12,\,13).
In this case, the chosen $P_s$ may intersect with the leaf's ranges,
and these ranges are divided and stored in both leaves.
It may cause the new leaf overflow (i.e., entry count exceeds the capacity and cannot be addressed by GC).
To resolve this, we select $P_{s}$ that satisfies:
(1) It is one of the two median points of all boundaries (as the number of boundaries is even);
(2) It is not the smallest or largest bound (to ensure the split is effective). 
If both medians meet the second requirement, we choose the smaller one.
We have proven that this selection strategy ensures: 
(1) no overflow in most cases;
(2) in other cases, only one leaf may overflow, which can always be addressed by splitting it again (please refer to \autoref{sec:split-proof} for the proof).

\begin{figure}
  \centering
  \includegraphics[width=\linewidth]{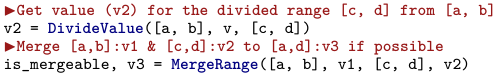}
\vspace{-8pt}
	\caption{\textbf{{\sys} user-provided functions.}}
	\label{fig:code:interface}
\end{figure}

When dividing a range that crosses $P_{s}$, this entry's value also needs to be divided.
The value is user-defined and may also represent a range  
(e.g., for {\org}'s {\ebs}, it is the location for a range of continuous physical blocks in DFS).  
Thus, {\plstree} needs users to register a function (\codeword{DivideValue} in \autoref{fig:code:interface}) to get the value corresponding to the divided ranges.
Specifically, it
returns the value for the divided ranges based on the original range, original value, and divided range.

\subsection{Workload-aware Merge and Resplit}
\label{sub:merge}
Unlike point indexes (e.g., B-tree), 
where merge only handles underfilled leaves caused by deletions,
{\plstree}'s merge operation also needs to reduce range fragmentation.
Range fragmentation may occur frequently because splits depend only on current ranges in the leaf and cannot foresee future workload, potentially causing mismatches between the leaf's range space and future insertions.
For example, although the split in \autoref{fig:normal_split} does not directly cause range fragmentation, inserting a new range [6,\,11] will span two leaves.
Thus, we propose a workload-aware merge/resplit mechanism to dynamically adjust the leaf's range space.

\begin{figure}[t]
  \centering
  \includegraphics[width=\linewidth]{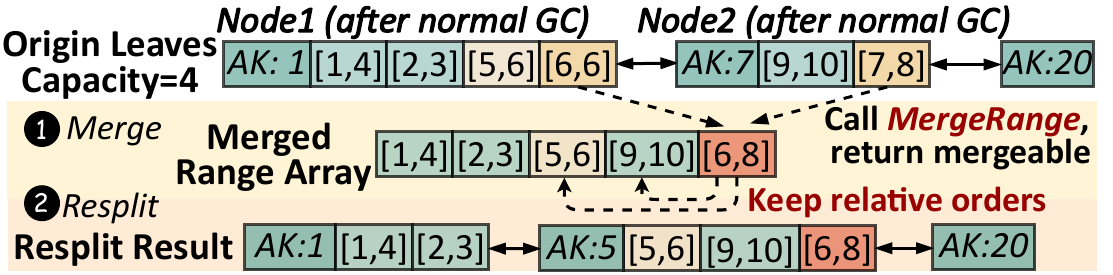}
  \vspace{-18pt}
  \caption{\textbf{Merge/resplit example.} \emph{AK} is anchor key.}%
  \label{fig:merge}
\end{figure}

To record the severity of range fragmentation for a period of time,
we use a leaf header field $N_{frag}$ to count the insertions of fragmented ranges\footnote{Fragmented ranges are subranges (except the first) after a range is divided.}.
When $N_{frag}$ exceeds a threshold,
we perform the merge/resplit procedure for this leaf and its left neighbor as shown in \autoref{fig:merge}. 
We first perform normal GC on both leaves to increase the likelihood of a successful merge.
Then, we aggregate two leaves' entries into a temporary merged array (\ding{182}).
If it fits in the leaf's capacity,
we update the left leaf using the merged array, remove the right leaf, and update the internal nodes;
otherwise,
we resplit the merged array (\ding{183}) to produce two new leaves.
The resplit is similar to the split, and it produces a more rational partitioning based on newer access patterns. 
When merging leaves, we also merge the ranges derived from the same user-written range, which can be identified by the user-provided function \codeword{MergeRange} (\autoref{fig:code:interface}).
\codeword{MergeRange} determines whether two entries can be merged based on their ranges and values.
If the two are mergeable (e.g, both their ranges and values are continuous), it returns the merged value for subsequent use.
For correct semantics, when merging ranges (e.g., [6,\,6] in Node1 and [7,\,8] in Node2), we preserve the relative order between the merged range and the leaf's other ranges
(e.g., order between [5,\,6]\,\&\,[6,\,8], [9,\,10]\,\&\,[6,\,8]).
Merging ranges reduces fragmentation and improves the success rate of merge.\footnote{More implementation details of RASK for EBS (e.g., functions in \autoref{fig:code:interface}, value format) and the default implementation of functions in \autoref{fig:code:interface} for general workloads are provided in \autoref{sec:more_design} for reference.}
 
%% file: impl.tex
\section{Operations and Concurrency Support}
\label{sec:rask-index-op}

\input{design/operation}

\input{design/concurrency}

%% file: design/operation.tex
\subsection{Operations}
\label{sub:op}

{\sys} supports three operations with the above designs:

\begin{figure}
  \centering
  \includegraphics[width=\linewidth]{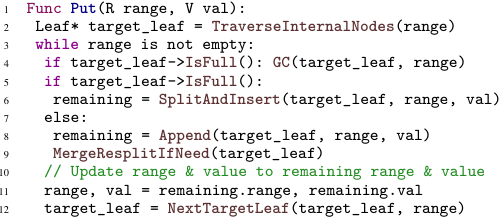}
	\caption{\textbf{Pseudocode of put operation.}}
	\label{fig:code:insert}
\end{figure}

\noindent
\textbf{Put\,(\emph{range, value}).}
As the pseudocode in \autoref{fig:code:insert},
\ding{182} we traverse internal nodes to locate the last leaf whose anchor key $\leq$ \emph{range} left bound, assigning it as the \emph{target leaf} for the new value (line\,2).
\ding{183} GC is performed if \emph{target leaf} is full (line\,4).
\ding{184} If the leaf is still full after GC, it is split and the new entry is inserted together (line\,6).
\ding{185} Otherwise, the new entry is appended directly (line\,8); then merge/resplit is performed if needed (line\,9).
We skip merge/resplit if a split occurs to avoid high write latency. 
\ding{186} As only the portion of the \emph{range} that fits in the \emph{target leaf}'s range space is inserted, the uninserted portion (i.e., fragmented range) is iteratively inserted into subsequent target leaves (line\,3-12). 
The next \emph{target leaf} is located through the doubly-linked leaf list (line\,12).

\noindent
\textbf{Delete\,(\emph{range}).}\,
Delete\,(\emph{range}) acts as Put\,(\emph{range, tombstone}) basically.
It inserts \emph{range} with a \emph{tombstone}
and increments $N_{frag}$ in the leaf's header.
During the normal GC, ranges with \emph{tombstone} are physically removed.
Specifically, 
we maintain an additional ordered list for ranges with \emph{tombstone} (i.e., \emph{deleted list}).
When handling a leaf entry during GC,
its overlapping portions with \emph{deleted list} are removed before excuting the GC logic.
During merge/resplit triggered by $N_{frag}$, the underfilled leaf is merged.
Note that reserving a value as \emph{tombstone} is a common and acceptable practice in indexes~\cite{wu2019wormhole,oneil1996log}.

\noindent
\textbf{Get\,(\emph{range}).}
\ding{182} It locates the last leaf with anchor key $\leq$ \emph{range} left bound (i.e., \emph{target leaf}) using the same method as \emph{Put}.
\ding{183} It traverses the doubly-linked leaf list 
to the last leaf with anchor key $\leq$ \emph{range} right bound.
\ding{184} It retrieves entries in each leaf via ablation-based search.
The \unfound is initialized with the intersection of the \emph{range} and the leaf's range space.
Entries with \emph{tombstone} are updated to the \unfound but excluded from the result.
Notably, we may only need the value of a subrange within a leaf's range (e.g., in \autoref{fig:search}, only [3,\,6] within [1,\,6] is required).
In this case, we use the \codeword{DivideValue} function (\autoref{fig:code:interface}) to retrieve the value of the subrange. 
Finally, it returns a list of values and their corresponding ranges.

%% file: design/concurrency.tex
\subsection{Concurrency Support}
\label{sub:concurrency}

{\sys} uses the standard optimistic lock-based concurrency control techniques~\cite{mao2012cache,bronson2010practical} (i.e., per-node write locks and version numbers).
Writes must acquire relevant lock(s) first.
Version numbers are incremented before and after the node's mutation.
Reads check version numbers before and after accessing the node, and retries if any version is changed or odd (indicating the node is being modified).
Since internal nodes directly employ ART's optimistic lock mechanism~\cite{leis2016art}, we focus on the concurrency safety of (1) inter-leaf interactions and (2) leaf and internal node interactions.

In {\sys}, after locating the \emph{target leaf} via read-only internal node traversal, there may be six concurrent operations: insert, delete, GC, split, merge/resplit, and read. Here we explain the safety of 
write-write concurrency and read-write concurrency through a case-by-case analysis.

\textbf{(1)\,Insert (also represents delete)}: 
To prevent concurrent leaf mutations (i.e., insert, delete, GC, split, merge/resplit),
it locks the target leaf(s).
For insertion across leaves, we use lock handover between leaves (i.e., lock next target node, then unlock current node) to prevent the disorder of cross-leaf concurrent updates.
Concurrent reads are safe and efficient because
reads can be treated as capturing a snapshot of append-only leaves, and inserts never trigger read retries. 

\textbf{(2)\,GC}: 
To prevent concurrent writes,
it locks the target leaf.
To ensure safe concurrent reads,
it updates $V_{GC}$ (4\,bits in leaf's header) on lock acquisition/release;
reads check $V_{GC}$ before\,and\,after accessing a leaf, and retry if necessary.

\textbf{(3)\,Split/merge/resplit}: 
\ding{182} To prevent concurrent writes,
splits lock the original and new leaves, 
and merge/resplits lock both leaves to be merged.
\ding{183} After the leaf is split/merged, it is inserted into (deleted from) the leaf linked-list.
Then the deleted nodes are marked as \emph{deleted} (i.e., a bit in the header) and reclaimed by epoch-based GC.
Same as B-tree, 
linked-list updates are safe without acquiring additional locks.
This is because the involved unlocked leaf must be the right neighbor of a locked leaf, whose \codeword{prev} pointer is protected by the lock in its locked left neighbor~\cite{mao2012cache}. 
\ding{184} After the linked-list update is done and the leaf locks are released, the internal nodes are updated. 
Before the internal nodes update is completed, 
the concurrent internal node traversal may locate a wrong \emph{target leaf}.
To ensure the correctness of \emph{target leaf},
after locating a leaf by internal node traversal, we traverse the doubly-linked leaf list to find the correct \emph{target leaf} as this list maintains the latest state.
\ding{185} To ensure safe concurrent reads,
split and merge/resplit employ $V_{split}$ and $V_{merge}$ (4\,bits in leaf's header) like GC
respectively.
Particularly, besides re-searching current leaf, read retries triggered by $V_{merge}$ also search the updated left leaf as values in the target range may have been moved there.
Retries trigggered by $V_{split}$ do not need to search the new right leaf, as rightward leaf traversal ensures any target values in it will be retrieved.
It is noteworthy that the split point selection is read-only and does not require synchronization with reads, thus $V_{split}$ only needs to be updated around the actual data movement.
While checking version numbers, reads also check the \emph{deleted} bit, and skip deleted nodes.

\begin{table}[t]
\centering
\caption{\textbf{Statistics of the evaluated traces.} The scale indicates the data volume used in our tests rather than the entire trace size.}
\vspace{-8pt}
\label{tab:dataset_spec}
\resizebox{0.9\columnwidth}{!}{
\begin{tabular}{lcccc}
\toprule
\textbf{Dataset} & \textbf{{\org}} & \textbf{Meta} & \textbf{Google} & \textbf{Tencent} \\ \midrule
\textbf{Scale}               & 1.5\,TB        & 150\,GB       & 92\,GB          & 588\,GB          \\
\textbf{Duration} & 1 week  & 3 year        & 3 month         & 10 days          \\ \bottomrule
\end{tabular}
}
\end{table}

\begin{table*}[t]
  \centering
  \caption{\textbf{Baseline selection.} For HOT and Cuckoo Trie, we do not implement {\rask} versions as their source code lacks the deletion interface. For PGM-index, we do not implement the Lazy version as it only supports numeric keys. As highlighted in bold, we choose Lazy B-tree, Lazy ART, original Wormhole, Eager Hydralist, and original PGM-index as subsequent baselines.}
  \label{tab:baseline_selection}
  \vspace{-8pt}
  \resizebox{\textwidth}{!}{
  \begin{tabular}{ccccccccccccccc}
    \toprule
  \multicolumn{1}{c}{\multirow{2}{*}{\textbf{Index Type}}} & \multicolumn{3}{c}{\textbf{B-tree}} & \multicolumn{3}{c}{\textbf{ART}}      & \multicolumn{3}{c}{\textbf{Wormhole}}  & \multicolumn{3}{c}{\textbf{HydraList}} & \multicolumn{2}{c}{\textbf{PGM-index}} \\ 
  \multicolumn{1}{c}{}     & Original   & Eager  & \textbf{Lazy} & Original & Eager  & \textbf{Lazy} & \textbf{Original}   & Eager    & Lazy     & Original    & \textbf{Eager} & Lazy & \textbf{Original} & Eager  \\ \hline
  \textbf{Throughput (Mop/s)} & 0.586   & 1.52  & \textbf{1.72}  & 1.51  & 1.28  & \textbf{1.53}   & \textbf{0.218}   & 0.124    & 0.190 & 0.0540 & \textbf{0.170}  & 0.050  & \textbf{0.725} & 0.003         \\
  \textbf{Memory (MB)}        & 470 & 50.2 & \textbf{48.6} & 1830 & 782 & \textbf{599} & \textbf{618} & 1820 & 1399 & 86.6  & \textbf{77.7} & 93.2  & \textbf{204}  & 147  \\ \bottomrule
  \end{tabular}
  }
\end{table*}

Based on the above techniques, {\sys}'s read operations can get a consistent view (i.e., snapshot) of each leaf, but cross-leaf reads may (1) partially read a user-written range or (2) miss earlier inserts while seeing later ones.
We argue that these issues are negligible in practice from both correctness and performance perspectives.
Regarding correctness, applications (e.g., EBS, DFS metadata service, flash cache) already can tolerate these issues because they currently rely on {\pask} indexes (e.g., Masstree~\cite{mao2012cache}, ART~\cite{binna2018hot}, Cuckoo-Trie~\cite{zeitak2021cuckoo}, and HydraList~\cite{mathew2020hydralist}), which exhibit the same issues when reading a range~\cite{zeitak2021cuckoo,binna2018hot,mao2012cache,mathew2020hydralist}.
Regarding performance, our breakdown of experiments in \autoref{sub:scalability-analysis} reveals that such inconsistent cross-leaf reads are exceedingly rare  (\textasciitilde0.0394\%), thus their impact on overall  performance and results is statistically negligible.

To further address the inconsistency of cross-leaf reads,
we can first snapshot all involved leaves while ensuring no concurrent writes, which we leave as future work.

For persistence, {\sys} is currently an in-memory index. Applications (e.g., EBS) are responsible for persistence.

%% file: eval.tex
\begin{figure*}[t]
  \centering
  \subfloat[Average throughput per {\vd}]{
      \includegraphics[width=0.497\textwidth]{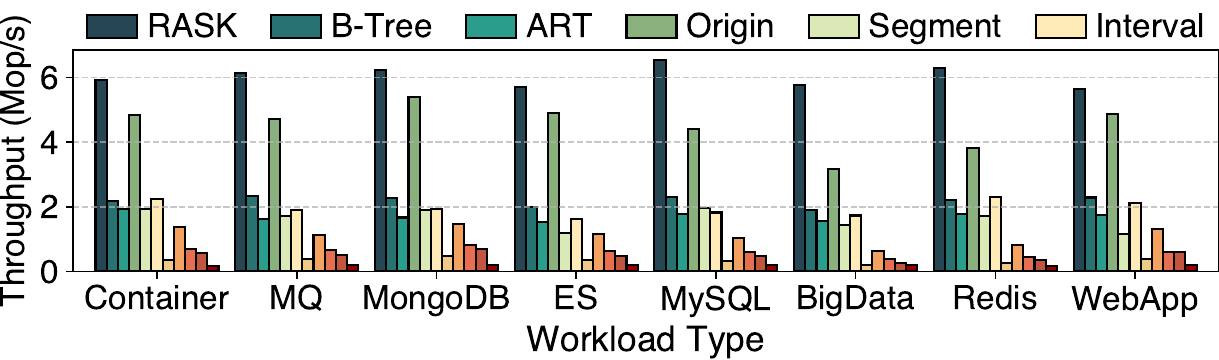}
      \vspace{-3pt}
      \label{fig:overall_throughput}}%
  \subfloat[Average memory footprint per {\vd}]{
          \includegraphics[width=0.484\textwidth]{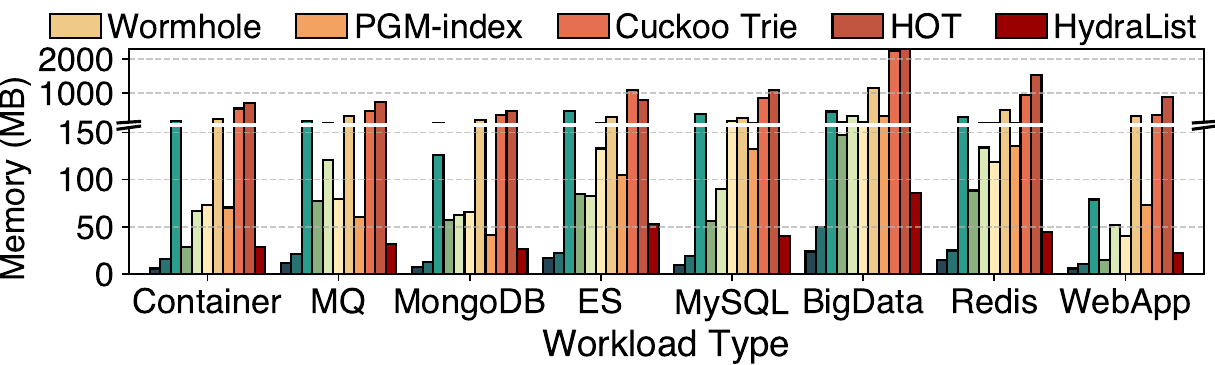}
          \vspace{-2pt}
          \label{fig:overall_memory}}%
   \vspace{-8pt}
  \caption{\textbf{Throughput and memory footprint of {\sys} and baselines on the Full Dataset.}}
  \label{fig:overall}
  \vspace{-5pt}
\end{figure*}

\begin{figure*}[t]
  \centering
  \includegraphics[width=\linewidth]{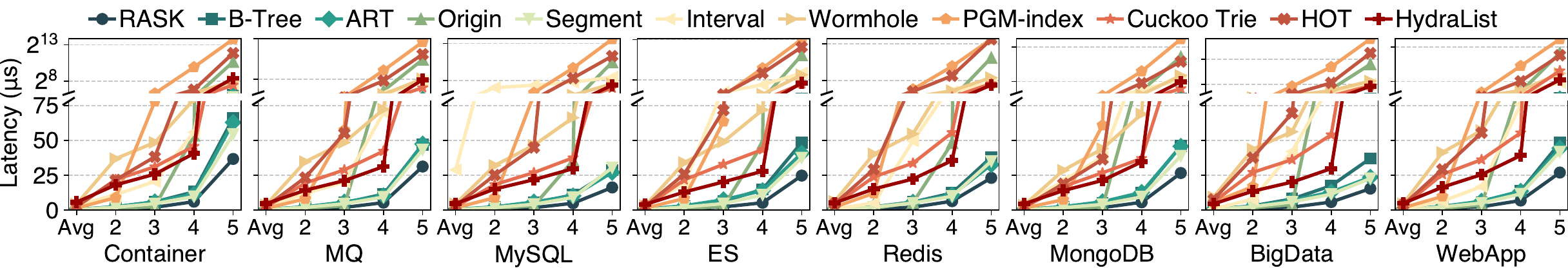}
  \vspace{-17pt}
  \caption{\textbf{Average and tail latency of {\sys} and baselines on the Full Dataset (per {\vd}).} In the x-axis tick labels, 2 represents P99, 3 represents P99.9, 4 represents P99.99, and 5 represents P99.999. All sub-figures share the same y-axis.}
  \label{fig:latency}
\end{figure*}

\begin{figure*}[t]
  \centering
  \begin{minipage}{0.4\linewidth}
      \centering
      \includegraphics[width=\linewidth]{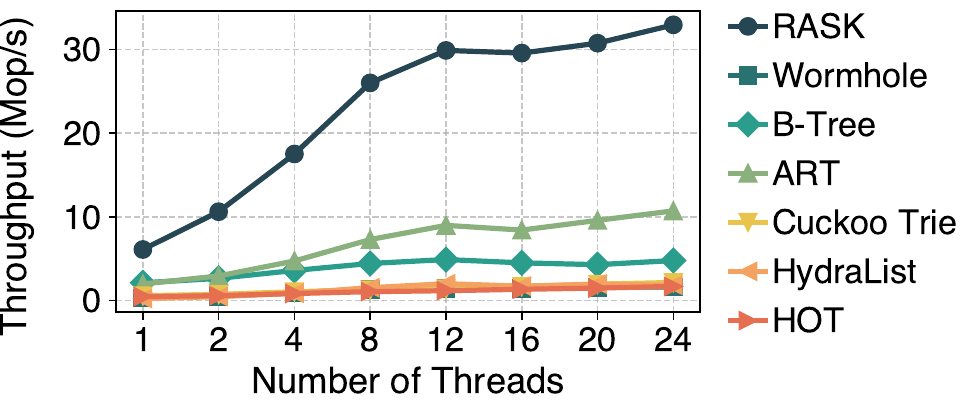}
      \vspace{-16pt}
      \caption{\textbf{Concurrency scalability.}}
      \label{fig:scalability}
      \vspace{-7pt}
  \end{minipage}
  \begin{minipage}{0.59\linewidth}
    \hfill
      \centering
    \begin{subfigure}[t]{0.475\linewidth} %
        \includegraphics[width=\linewidth]{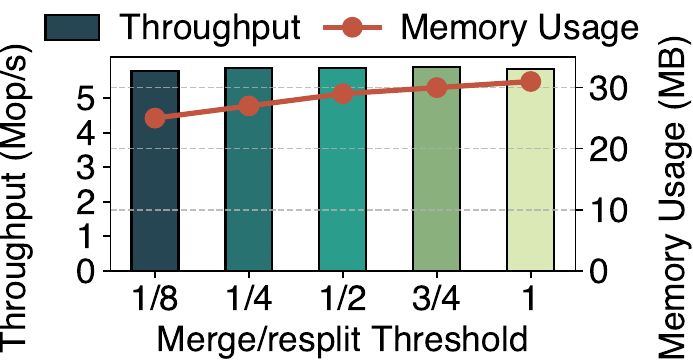}
        \vspace{-15pt}
        \caption{Merge/resplit trigger threshold}
        \label{fig:merge_threshold}
    \end{subfigure}
    \hfill
    \begin{subfigure}[t]{0.475\linewidth}
        \includegraphics[width=\linewidth]{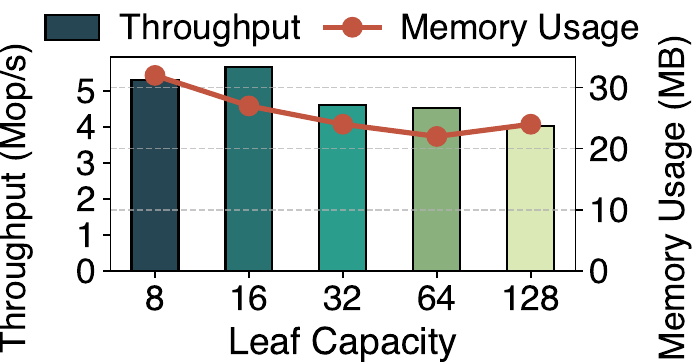}
        \vspace{-15pt}
        \caption{Leaf capacity}
        \label{fig:leaf_node_count}
    \end{subfigure}
    \vspace{-7pt}
    \captionsetup{type=figure}
    \caption{\textbf{Impact of {\sys}'s parameters on the {\sampled}.}}
    \label{fig:param_sensitivity}
    \vspace{-7pt}
  \end{minipage}%
\end{figure*}

\begin{figure*}[t]
  \centering
  \subfloat[Varied write range length]{
      \includegraphics[width=0.49\textwidth]{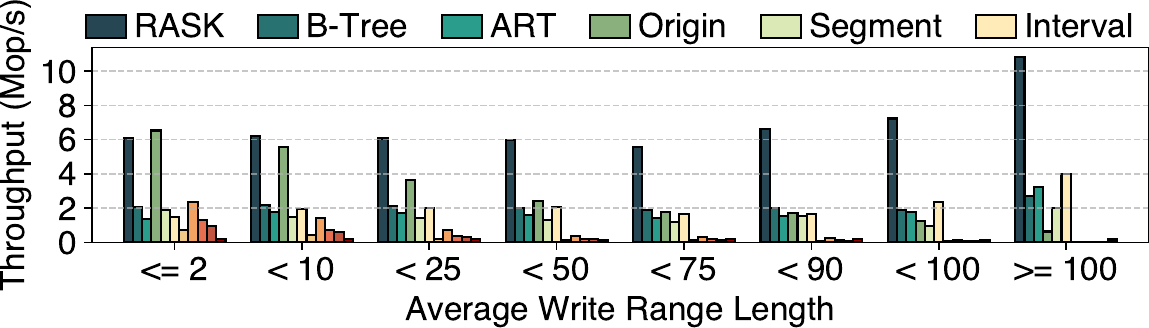}
      \vspace{-3pt}
      \label{fig:varied_write_length}}
  \hfill
  \subfloat[Varied read/write ratio]{
      \includegraphics[width=0.485\textwidth]{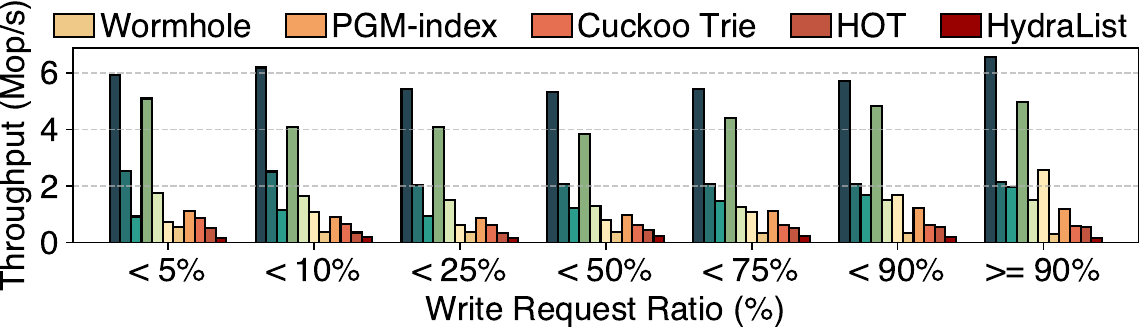}
      \vspace{-2.5pt}
      \label{fig:varied_write_ratio}}
   \vspace{-7pt}
  \caption{\textbf{Performance comparison between {\sys} and baselines under different workload characteristics on the {\full}.}}
  \label{fig:vary_workload}
  \vspace{-5pt}
\end{figure*}

\begin{figure*}
  \centering
  \includegraphics[width=\linewidth]{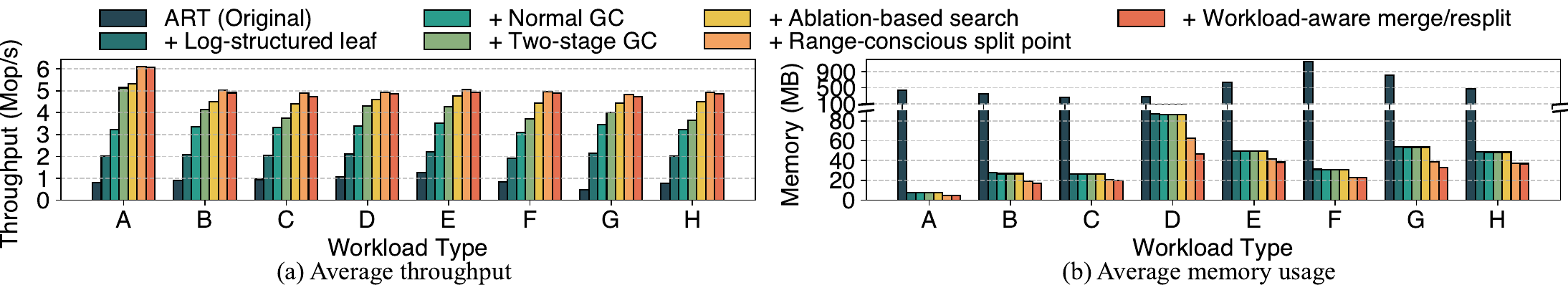}
  \vspace{-20pt}
  \caption{\textbf{The factor analysis for techniques in {\sys} on the {\sampled}.} Workloads are: \emph{Container}\,(A), \emph{MQ}\,(B), \emph{MySQL}\,(C), \emph{ES}\,(D), \emph{Redis}\,(E), \emph{MongoDB}\,(F),
      \emph{BigData}\,(G), \emph{WebApp}\,(H).}
      \label{fig:breakdown}
  \vspace{-5pt}
\end{figure*}

\begin{figure}
  \centering
  \includegraphics[width=0.95\columnwidth]{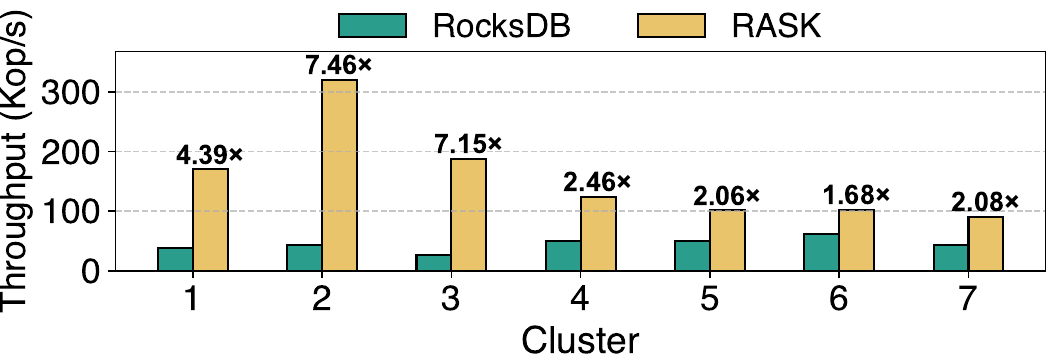}
  \vspace{-7pt}
  \caption{\textbf{Comparison on Meta traces.}} 
  \vspace{-5pt}
  \label{fig:tectonic}
\end{figure}

\begin{figure*}[t]
  \centering
  \begin{minipage}{0.49\linewidth}
    \centering
    \includegraphics[width=\linewidth]{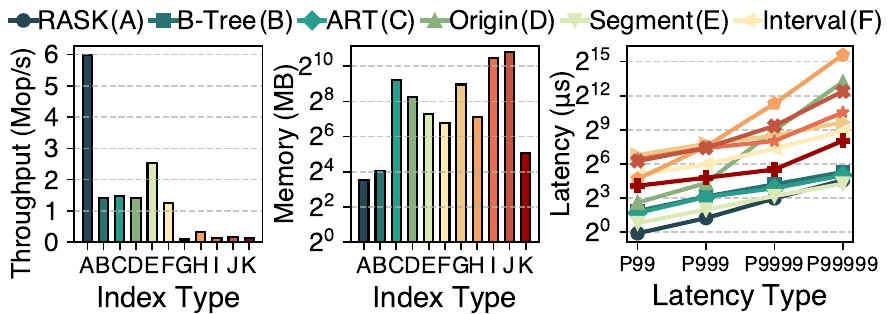}
    \vspace{-17pt}
    \caption{\textbf{Comparison on Tencent traces.}}
    \label{fig:tencent}
    \vspace{-5pt}
  \end{minipage}
\hfill
  \begin{minipage}{0.49\linewidth}
      \centering
    \includegraphics[width=\linewidth]{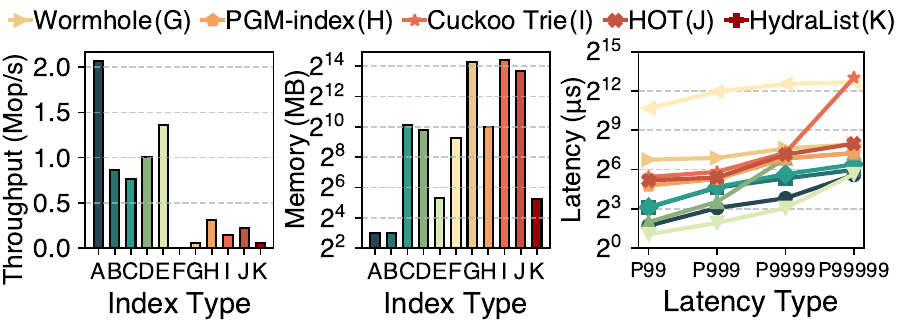}
    \vspace{-17pt}
    \caption{\textbf{Comparison on Google traces.}}
    \label{fig:google}
    \vspace{-5pt}
  \end{minipage}%
\end{figure*}

\section{Evaluation}%
\label{sec:eval}

We evaluate {\sys} from the following perspectives:
\begin{itemize}[leftmargin=1em,topsep=-3pt]
  \item Does {\sys} perform well with different workloads? (\autoref{sub:overall-performance})
  \item How is the concurrency scalability of {\sys}? (\autoref{sub:scalability-analysis})
  \item How do workload characteristics and parameters affect {\sys}’s performance? (\autoref{sub:sensitivity-analysis})
  \item How do {\sys}'s techniques contribute? (\autoref{sub:performance-breakdown})
  \item Can other scenarios benefit from {\sys}? (\autoref{sub:case-study})
\end{itemize}

\subsection{Experimental Setup}
\label{sec:eval-setup}

\myparagraph{Setup.}
We run experiments in \autoref{sub:case-study} on a server with 2 Intel\textsuperscript{\textregistered} Xeon\textsuperscript{\textregistered} Gold 5317 CPUs (3.00\,GHz, 12 cores), 188\,GB DRAM and 7\,TB NVMe SSD.
Other experiments are run on a server with 1 Intel\textsuperscript{\textregistered} Xeon\textsuperscript{\textregistered} Platinum 8369B CPU (2.70\,GHz, 24\,cores) and 96\,GB DRAM.
All experiments except those in \autoref{sub:scalability-analysis} and case\,3 in \autoref{sub:case-study} are single-threaded as the internal logic of the cloud block store is single-threaded.

\myparagraph{Parameters.}
By default, we configure {\sys}'s parameters as follows:
(1) The leaf's capacity is 16;
(2) The merge/resplit trigger threshold is 4 (i.e., $\frac{1}{4}$ of the leaf size).
We also evaluate {\sys} across varied parameter settings in \autoref{sub:sys-parameters}.

\myparagraph{Benchmarks.}
Our primary benchmark is EBS traces of 1.8\,k VDs from 4 clusters (\autoref{sec:trace}) at {\org}.
We use post-I/O compaction (\autoref{sub:write-write-correlation}) results of these traces as \emph{Full Dataset}.
We sample 100 {\vd}s to form \emph{Sampled Dataset}, covering all workloads at all load levels.
Since an experiment on {\full} takes at least 5--7 days,  
we only use \full for overall experiments (\autoref{sub:overall-performance}, \autoref{sub:workload-characteristics}) and use \sampled for the others.
We also use traces from Tencent, Meta, and Google to show {\sys}'s effectiveness 
in other vendor's EBS and various scenarios (e.g., flash cache, DFS metadata service).

\myparagraph{Trace specifications.} The traces from {\org} EBS contain a sequence of I/O requests with LBA, length (in 4\,KB blocks), type (read/write), and timestamp.
We replay these requests in chronological order, with original time intervals omitted. This is a common evaluation practice~\cite{wang2024baleen,yang2023fifo}, as omitting time intervals does not affect the index's memory usage or single-threaded performance, and better reveals its behavior under multi-threaded contention. 
Traces from other vendors are in similar formats and are replayed similarly.
We further provide more statistics about these traces in \autoref{tab:dataset_spec}.

\myparagraph{Baselines.}
As a general-purpose index optimized for range-write heavy workloads,
{\sys} can benefit any apps with such characteristics, no matter which index it currently uses.
To verify this,
we compare {\sys} with 9 SOTA ordered indexes\footnote{HINT~\cite{chris2022hint} is a SOTA range-aware index, we do not compare with it as its open-sourced implementation does not support updates.}: Cuckoo Trie~\cite{zeitak2021cuckoo}, HydraList~\cite{mathew2020hydralist}, Wormhole~\cite{wu2019wormhole}, HOT~\cite{binna2018hot}, PGM-index~\cite{ferragina2020pgm}, thread-safe STX B-tree~\cite{xu2023bp,bingmann_stx_btree}, (ROWEX) ART~\cite{leis2016art}, segment tree~\cite{bently1980multidimensional}, and interval tree~\cite{bentlay1979algorithms,cormen2009introduction}.
We also compare with the EBS-index (\autoref{sub:elastic-block-storage}), refered as {\origin}.
For EBS-index, we use its default configuration 
(converting to SSTable when MemTable's size reaches 128\,k LBA).

For the seven point indexes in baselines,
we implement their Eager and Lazy {\rask} variants (\autoref{sub:not_off_the_shelf}) if possible for fairer comparison.
\autoref{tab:baseline_selection} shows their throughput and memory usage on {\sampled}.
We select their best-performing and most memory-efficient versions for subsequent experiments.

\subsection{Overall Performance}
\label{sub:overall-performance}

\myparagraph{Throughput.}
\autoref{fig:overall_throughput} shows {\sys}'s throughput reaches 2.76--37.8$\times$ that of 9 SOTA ordered indexes.
Their poor performance stems from: (1) handling range overlaps per write and read (B-tree, ART, Hydralist, Wormhole), (2) splitting a range operation into multiple point operations (HOT, Cuckoo Trie), (3) frequent SMOs (PGM-index), and (4) accumulating obsolete ranges (segment tree, interval tree).  
The throughput of {\sys} is 1.15--1.82$\times$ that of Origin, proving {\rask}'s advantage over per-LBA updates.
{\sys}'s log-structured leaf greatly reduces the frequency of GC, while two-stage GC and ablation-based search speed up overlap handling.

\myparagraph{Memory footprint.}
As \autoref{fig:overall_memory} shows, {\sys} uses only 1.15--54.7\% of baselines' memory.
Notably, it requires only \textasciitilde19.9\% of {\origin}'s memory, 
showing that {\sys} can greatly reduce the memory footprint of EBS-index.
Moreover, 
{\sys} uses only 5.31--20.5\% of the memory required by the segment and interval tree, proving its memory efficiency from removing covered ranges.
For baselines, 
Cuckoo Trie and HOT consume the most memory as they are {\pask} indexes requiring more entries.
The memory overhead of the segment and interval tree mainly stems from the obsolete ranges.
B-tree is the most memory-efficient baseline due to its multi-entry leaves.
Trie-based indexes (Cuckoo Trie, ART, HOT) consume more memory due to single-entry leaves.
{\sys} and HydraList mitigate this by packing multiple entries per leaf.

\myparagraph{Latency.}
We measure both average and tail latency distributions (P99--P99.999) for {\sys} and baselines.
As shown in \autoref{fig:latency}, {\sys} achieves lower average and tail latency across all workloads, reducing P99 latency by 23.9--97.6\% and P99.999 by 34.2--99.7\% vs. baselines.
It confirms the efficiency of log-structured design, two-stage GC, and range-tailored split/merge.
Notably, {\sys} reduces P99.99/P99.999 latency by 90.9\%/98.8\% vs. Origin by avoiding write stalls from {\indexmap}'s LSM-like structure.
For other baselines, their high tail latency is mainly due to the frequent SMOs from handling range overlaps or using multiple point writes to achieve range updates.
As a learned index, PGM-index exhibits the worst tail latency
among baselines 
because of the overhead from model retraining and cascading updates.

\subsection{Scalability Analysis}
\label{sub:scalability-analysis}

\autoref{fig:scalability} compares the multi-threaded performance of {\sys} with other thread-safe baselines.
Note that {\org}'s EBS traces are inherently single-threaded for each {\vd}.
Therefore, we create multi-threaded workloads by distributing operations of a single {\vd} trace evenly across multiple threads while maintaining their original order. 
{\sys}'s throughput reaches 3.08--21.5\,$\times$ that of the baselines at 24 threads.
This is because baselines use read-optimized concurrency control and in-place updates.
Range writes and handling range overlaps cause longer write time,
leading to higher contention and lower throughput.
In contrast, {\sys}'s log-structured leaf reduces lock duration 
for most writes, 
and 
version numbers are modified only around node mutations, reducing read-write conflicts.
Thus, {\sys} scales well with 1--12 threads.
Beyond 12 threads, performance growth slows as some write-heavy, highly skewed traces increase contention. 
Our further analysis reveals that {\sys} still scales well for traces with more reads and less skewed writes at 12--24 threads.

We also measure the latency of {\sys} and baselines under multi-threading.
Results show that as thread count increases, {\sys} outperforms baselines more significantly in both average and tail latency.
At 24 threads, {\sys} reduces average latency by 85.9--98.3\% and tail latency by 82.3--99.9\% vs. baselines.
It proves that {\sys}'s optimistic concurrency control is efficient, and GC/SMOs are not tail-latency bottlenecks.
Particularly, split/merge merely blocks concurrent writes (< 0.01\% cases), indicating negligible contention.

\subsection{Sensitivity Analysis}
\label{sub:sensitivity-analysis}

In this section, we evaluate the sensitivity of {\sys} from two perspectives:
(1) {\sys}'s parameters (\autoref{sub:sys-parameters}), and (2) workload characteristics (\autoref{sub:workload-characteristics}).

\subsubsection{Impact of {\sys}'s parameters}
\label{sub:sys-parameters}

\noindent
\textbf{Impact of merge trigger threshold.}
\autoref{fig:param_sensitivity}(a)
shows the throughput and memory usage of {\sys} when merge/resplit trigger threshold is $\frac{1}{8}$, $\frac{1}{4}$, $\frac{1}{2}$, $\frac{3}{4}$, and 1 of the leaf size.
A lower threshold increases merge/resplit frequency, saving memory (\textasciitilde24.0\%) but slightly lowering throughput (\textasciitilde1.67\%).

\noindent
\textbf{Impact of leaf capacity.}
\autoref{fig:param_sensitivity}(b)
shows {\sys}'s performance trend as the leaf capacity increases from 8 to 128 entries:
the peak throughput occurs at 16 entries.
Fewer entries increase GC frequency, hurting throughput. 
Meanwhile, more entries raise the overlap-handling cost, also degrading performance.
Memory usage drops 31.3\% from 8 to 64 entries as fewer nodes lower metadata costs,
but rises 9.09\% at 128 entries due to greater memory waste from idle entries.

\subsubsection{Impact of workload characteristics}
\label{sub:workload-characteristics}

\noindent
\textbf{Impact of range length.}
\autoref{fig:varied_write_length} shows that {\sys}'s advantage scales with longer ranges.
For short ranges (average length\,<\,10), it achieves 1.84--12.71$\times$ higher throughput than 9 ordered indexes and 11.1\% higher than {\origin}.
Under worst-case sparse writes (average length\,$\leq$\,2), {\sys} outperforms 9 ordered indexes by at least 1.56$\times$. 
It only slightly underperforms {\origin} by 6.64\%, owing to {\origin}'s efficient $O(1)$ updates for small writes.  
This indicates that tasks with range length\,>\,2 can typically benefit from {\sys}.
When the average range length $\geq$ 100, {\sys} outperforms {\origin} by 16.10$\times$ and exceeds other 9 baselines by 17.2--312.97$\times$.

\noindent
\textbf{Impact of read/write ratio.}
\autoref{fig:varied_write_ratio} shows that {\sys} performs well across all read/write ratios.
It outperforms baselines by 16.0\%--37.3$\times$ for read-heavy workloads (write ratio < 5\%), 39.1\%--38.7$\times$ for write-heavy workloads (write ratio $\ge$ 90\%).
As GC frequency correlates with write ratio, this also proves {\sys}'s strong performance under varying GC frequencies.
This stems from:
(1) The log-structured leaf speeds up writes while enabling early termination for efficient reads;
(2) The ablation-based search enhances query efficiency.

\subsection{Breakdown Analysis}
\label{sub:performance-breakdown}
\
\myparagraph{Contributions of techniques.}
We apply each technique in {\sys} one by one to the original ART with optimistic locking.
\autoref{fig:breakdown} shows how each technique contributes to the overall performance and memory usage.

\textbf{+\,Log-structured leaf.} 
We scale the leaf size from single-entry to multi-entry, adapt the log-structured design, and support {\rask}.
It improves throughput by 1.50$\times$ and reduces memory usage by 90.3\% compared to the original ART.
The total frequency of GC relative to write operations is 5.08\%, indicating that 
the log-structured design effectively circumvents the overhead of handling range overlaps for most writes.

\textbf{+\,Normal GC.}
This step improves the throughput by 70.6\% on average.
This gain is primarily due to normal GC maintaining the {\nonoverlap}, which helps efficiently identify old ranges covered by multiple new ranges for batch deletion.

\textbf{+\,Two-stage GC.}
In this step, in addition to normal GC, we implement lightweight GC, forming a two-stage GC mechanism.
This step improves average throughput by 24.1\% compared to the normal GC alone, owing to the lightweight GC's ability to quickly free occupied slots, thereby reducing blocking time for incoming writes.
The average effectiveness probability of lightweight GC is 59.1\%.

\textbf{+\,Ablation-based search.}
This step improves the average throughput by 12.6\% even though most traces are write-heavy, proving the effectiveness of the ablation-based search.

\textbf{+\,Range-conscious split.}
This step improves the average throughput by 7.56\% and cuts memory usage by 26.0\%, showing this technique effectively identifies suitable split points.

\textbf{+\,Workload-aware merge and resplit.}
This step reduces the average memory usage by 7.70\%, 
despite it incurs a slight performance overhead (1.90\%) due to merge/resplit checks.
In practice, the merge/resplit only occurs when necessary, keeping its overhead low (avg. frequency: 0.87\% of writes).

\myparagraph{Breakdown of range-conscious split.}
We further analyze the quality of split points obtained by range-conscious split.
Across all workloads, 84.3\% of splits do not divide any range, effectively avoiding range fragmentation.
Less than 0.01\% splits require a second split---making the overhead of second splits negligible.
The average difference in entry counts between the two new leaves is less than 2.34,
showing that the range-conscious split can achieve relatively balanced splits.

\subsection{General Applicability Analysis}
\label{sub:case-study}

We apply {\sys} to the following three scenarios to show that {\sys} is not only effective for {\org}'s EBS, but also generalizable to other scenarios with extensive range writes.

\noindent
\textbf{{Case 1: EBS of other vendors (Tencent trace).}}
To verify {\sys}'s applicability beyond {\org}, we\,evaluate\,it\,on Tencent's EBS traces (thousands of {\vd}s over ten days).
As \autoref{fig:tencent} shows, 
{\sys} achieves 2.35--49.21$\times$ throughput, 27.4--99.3\% lower memory usage, and 46.4--98.8\% lower tail latency vs. baselines, 
as range-write is a widespread pattern in EBS.
This proves {\sys}'s cross-vendor effectiveness.

\myparagraph{Case 2: DFS metadata service (Meta trace).}
Meta's exabyte-scale DFS (Tectonic) uses RocksDB~\cite{zippydb2015,dong2021rocksdb} for metadata service, including a block-to-file mapping. 
To validate {\sys}'s effectiveness in this scenario, we replace RocksDB's MemTable with {\sys} and evaluate it with Tectonic's 3-year traces (from 7 clusters).
For RocksDB, we use its default configuration, including various parameters and concurrency control mechanisms (i.e., MVCC).
\autoref{fig:tectonic} shows
{\sys} achieves up to 7.46$\times$ the throughput of the original RocksDB
since {\sys} is more memory-efficient than skiplist, allowing more entries to stay in memory and boosting query performance.

\myparagraph{Case 3: Flash cache index (Google trace).}
Given Google's storage clusters using flash cache~\cite{phothilimthana2024thesios}, 
we simulate {\sys}'s effectiveness in flash cache scenarios using 3-month traces from 3 Google clusters.
\autoref{fig:google} shows that {\sys} achieves 1.52--37.52$\times$ higher throughput, reduces memory usage by 3.2--99.9\%, and cuts tail latency by 4.2--99.4\% vs. baselines.
The only exception is that segment tree's tail latency is lower than {\sys}, as it  avoids the overhead of removing covered ranges---which has minor impact under light load---but incurs high memory overhead.

%% file: related.tex
\section{Related Work}%
\label{sec:related}

\myparagraph{Memory-efficient indexes in block devices.}
In SSDs, the Flash Translation Layer (FTL) handles logical-to-physical (L2P) mapping~\cite{GalToledo2005}, 
which is ideally cached in memory.
Various device-side FTL designs have explored reducing the memory footprint by leveraging I/O patterns~\cite{zhou2015efficient,gupta2009dftl,lee2007log,lee2008last,im2025appl}, with
some exploiting write sequentiality~\cite{jiang2011sftl,sun2023leaftl}.
However,
to meet SSD requirements,
these designs still operate at the LBA granularity rather than range, thus incurring additional conversion and operational overhead.
On the other hand, host-side FTLs for open-channel SSDs~\cite{bityutskiy2005jffs3,ouyang2014sdf,zhang2017flashkv}, Zoned Namespaces (ZNS)~\cite{bjorling2021zns,Kyuhwa2021zns+,oh2023zenfs+,stavrinos2021dont}, and Flexible Data Placement (FDP)~\cite{manzanares2023fdp,desai2021smartftl} reduce memory overhead by degrading page-level mapping to zone/reclaim unit-level, thus reducing mapping entries.
However, this fixed range size (e.g., zone size) limits flexibility and imposes more strict requirements on applications.
{\sys} natively supports variable-length range indexing, making it more flexible and applicable for cloud block storage systems.

\myparagraph{B-tree/Trie-based in-memory indexes.}
Current SOTA in-memory ordered indexes are primarily based on B-trees~\cite{xu2023bp,wang2018bw,awad2019engineering,cja2023blink}, tries~\cite{zeitak2021cuckoo,binna2018hot,leis2013adaptive}, or their hybrids~\cite{mao2012cache,kim2010fast}.
HydraList~\cite{mathew2020hydralist} and Wormhole~\cite{wu2019wormhole} use trie-like internal nodes for fast leaf access and B-tree-like leaves for efficient range queries and memory efficiency.
While {\plstree} also uses trie-style internal nodes and B-tree-style leaves, it is further designed for indexing ranges rather than individual objects.

\myparagraph{Memory-efficient persistent KV stores.}
Persistent KV stores (e.g., LSM-tree) require an in-memory index and/or cache for faster data access.
Many works have improved the performance of KV stores by optimizing cache~\cite{lu2016wisckey,dai2020bourbon,yang2020leaper,luo2020breaking,wu2020ackey} and indexing~\cite{balmau2017flodb,bortnikov2018accordion,oana2017triad,lepers2019kvell,Brinkmann2025hln}.
{\sys} can serve as the in-memory component for persistent KV stores, enhancing their memory efficiency.

%% file: concl.tex
\section{Conclusion}%
\label{sec:concl}

We propose {\sys}, a memory-efficient and high-performance index that natively supports range-as-a-key.
We employ several techniques to address the challenges of range overlap and range fragmentation.
Evaluation on four industry traces shows {\sys}'s advantages compared to ten SOTA indexes.

%% file: ack.tex
\section*{Acknowledgments}
We thank our shepherd Youjip Won and the anonymous reviewers for their insightful comments and feedback.
We are sincerely grateful to Jingkai He for his valuable support throughout this work, and to Yaheng Song and Shizhuo Sun for their dedicated efforts in collecting and processing the trace data.
We also thank Qiuping Wang, Jifei Yi, Jingyao Zeng, and Tong Xin for their helpful suggestions.
This work is supported in part by the National Natural Science Foundation of China (No. 62132014), the Fundamental Research Funds for the Central Universities, the Fundamental and Interdisciplinary Disciplines Breakthrough Plan of the Ministry of Education of China
(JYB2025XDXM113), and the Alibaba ARF/AIR program.
Corresponding authors: Mingkai Dong (\url{mingkaidong@sjtu.edu.cn}) and Erci Xu (\url{xjostep90@gmail.com}).

%% file: appendix/cu_exp.tex
\section{{\seqwritestream} with Different Window Sizes}
\label{sec:cu-exp}

\begin{figure}[H]
  \centering 
  \includegraphics[width=0.87\linewidth]{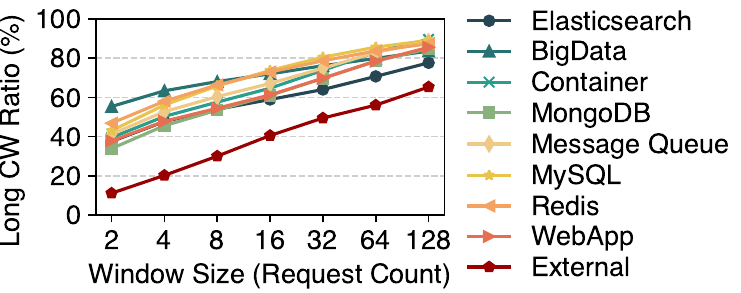}
  \vspace{-5pt}
  \caption{\textbf{Long {\seqwritestream} ratio with different window sizes.}}
  \label{fig:seqwritestream}
  \vspace{-10pt}
\end{figure}

To verify the prevalence of {\seqwritestream} under different window sizes, 
we measure the long {\seqwritestream} ratio across observation windows ranging from 2 to 128 write requests.
As shown in \autoref{fig:seqwritestream}, the long {\seqwritestream} ratio is 33.9--55.4\% when the window size is 2, and increases to 77.6--89.5\% when the window size is 128.
Tencent's trace (\emph{External} in \autoref{fig:seqwritestream}) shows a similar pattern: the long {\seqwritestream} ratio rises from 11.2\% to 65.3\% as window size increases from 2 to 128.
This indicates the widespread presence of {\seqwritestream}s.
Moreover, a larger window captures more {\seqwritestream}s because it allows interrupted write streams (from multi-app contention) to be observed as consecutive sequences.

%% file: appendix/cu_alignment.tex
\section{CU Alignment}

\subsection{Adaptive Compression Scheme}
\label{sub:cu-align}

Given the write-read correlation, we propose an adaptive compression scheme:  
{\seqwritestream}s exceeding 4 blocks are compressed in units of {\seqwritestream}s to reduce indexing overhead, 
while shorter {\seqwritestream}s use the 4-block CU to maintain the compression ratio.

\myparagraph{Continuous read optomization.}
To optimize continuous reads, we leverage a common feature of compression schemes\\~\cite{lz42025,zstd2025,google_snappy}---incremental decompression capability (i.e., resuming decompression from the previous stopping point)---by caching the last partially read CU in a \emph{decompression cache}.
This allows the next read resume decompression where the last one ended, avoiding redundant network I/O, disk I/O, and re-decompression from the CU start.

\begin{figure}[t]
  \centering
  \includegraphics[width=0.9\linewidth]{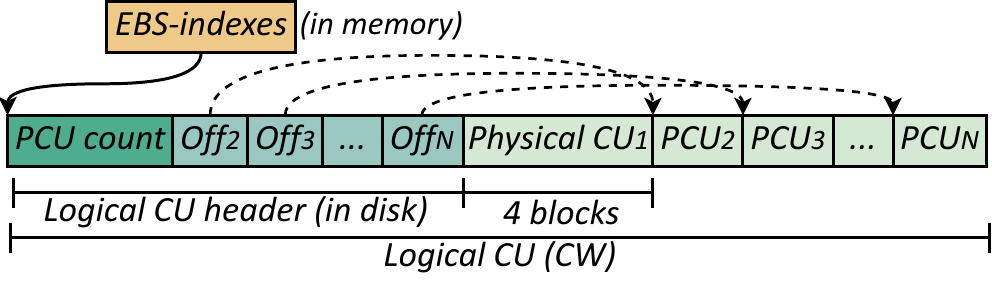}
  \vspace{-5pt}
  \caption{\textbf{Logical-physical CU scheme.} PCU refers to physical CU.}
  \label{fig:dual-layer-cu}
  \vspace{-5pt}
\end{figure}

\myparagraph{Long {\seqwritestream} optimization.}
To prevent long I/O and decompression times when reading from the middle of long {\seqwritestream}s (> 16 blocks), as shown in \autoref{fig:dual-layer-cu}, we design the \emph{logical-physical compression scheme}:
(1) Logical CU\,: the {\seqwritestream} serves as a logical CU and its offset information is indexed by the EBS-index.
(2) Physical CU\,: the {\seqwritestream} is subdivided into physical CUs of 4 blocks each. The physical CU's offset information is stored in the logical CU's header (not cached).
The overhead of the logical CU header is negligible since it only adds about 0.097\% storage and read cost per physical CU.

\myparagraph{Read procedure.}
(1) For original CU (i.e., CU $=$ 4), we read it as before.
(2) For CU $>$ 4 and $\leq$ 16, we fetch data based on the smaller size between the needed blocks' uncompressed size (i.e., read offset within CU + read length) and the CU size.
For the fetched data, we only decompress up to the position required by the read request. 
If the fetched data is not fully consumed, we cache the remaining data and decompression state in the decompression cache.
If the next request is read and hits the cache, decompression resumes from where it left off;
otherwise, we evict the cached data.
(3) For CU\,$>$\,16,
if read offset within the CU \,$\leq$\,16, we apply the procedure as in (2) (while also reading the logical CU header);
otherwise, we first read and parse the logical CU header to locate the target physical CUs, then fetch and decompress these physical CUs.

\subsection{Evaluation through Traces}
\label{sub:trace-eval}

We evaluate the potential benefit and overhead of the adaptive compression scheme through simulation and theoretical analysis on {\org}'s production traces.

\begin{figure}[t]
  \centering
  \includegraphics[width=0.9\linewidth]{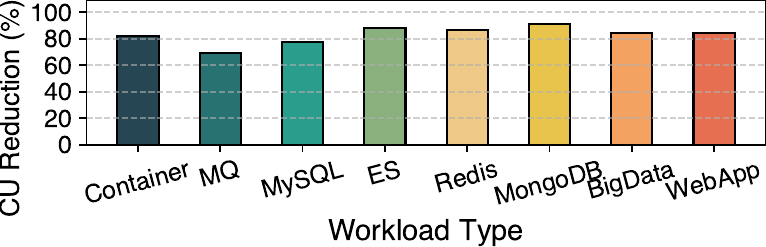}
  \vspace{-5pt}
  \caption{\textbf{Percentage of {\compressunit} count reduction after adopting {\compressunit} alignment.}}
  \label{fig:decrease-cu}
  \vspace{-10pt}
\end{figure}

\subsubsection{Potential Benefit}
We simulate the number of {\compressunit}s requiring indexing under {\compressunit} alignment by replaying traces, then compare it with the original scheme. 
As shown in \autoref{fig:decrease-cu}, the number of {\compressunit}s to be indexed reduces by 69.1--91.1\% across eight workloads.

\subsubsection{Memory Overhead}

The memory overhead of the adaptive compression scheme mainly comes from the decompression cache.
Since the maximum size of the decompression cache is 16 blocks (\autoref{sub:cu-align}), which is $\frac{1}{8}$ of the {\segmentcache}.
Additionally, in production environments, the {\segmentcache} occupies about 10\% of the cluster's memory.
Thus, the decompression cache theoretically adds at most 1.25\% memory overhead to the cluster.

In practice, the decompression cache's memory footprint is significantly lower than the theoretical upper limit.
This is because the decompression cache is allocated on demand (i.e., only for potential continuous reads) and released immediately if there are no continuous reads.
By replaying I/O requests from 1.8\,k {\vd}s across four clusters chronologically, the peak decompression cache memory footprint is only 0.012\%, 0.013\%, 0.023\%, and 0.023\% of total memory, respectively.

\subsubsection{Performance Overhead}

The performance overhead of the adaptive compression scheme mainly comes from two aspects:
(1) increased disk I/O and network I/O due to fetching larger {\compressunit}s;
(2) increased decompression time 
since the read may start further from the {\compressunit} beginning than before. 

We define the original scheme fetches $N_{ori}$ bytes of data and the read offset from the {\compressunit} beginning as $D_{ori}$. 
The adaptive compression scheme fetches $N_{new}$ bytes of data and the read offset from the {\compressunit} beginning as $D_{new}$.
Moreover, we define the disk I/O speed as $V_{disk}$, network I/O speed as $V_{net}$, decompression speed as $V_{dec}$, the read request length as $R_{size}$, the {\compressunit} size as ${CU}_{size}$, and the logical {\compressunit} header size as $H_{size}$.

The additional access latency introduced by the adaptive compression scheme is:
\begin{equation*}
  \Delta T = (N_{new} - N_{ori}) \times (V_{net} + V_{disk}) + (D_{new} - D_{ori}) \times V_{dec} 
\end{equation*}

Specifically, the value of $N_{new}$ in the three cases of {\compressunit} size is as follows:
\begin{itemize}[leftmargin=1em,topsep=-1pt]
  \item Original CU: $N_{new} = N_{ori}$;
  \item CU $>$ 4 and $\leq$ 16: $N_{new} = \min(CU_{size}, D_{new} + R_{size})$;
  \item CU $>$ 16:
    \begin{itemize}[leftmargin=1em,topsep=-2pt]
      \item If $D_{new} < 16$, $N_{new} = \min(CU_{size}, D_{new} + R_{size})$;
      \item If $D_{new} \geq 16$, $N_{new} = H_{size} + N_{need}$, where $N_{need}$ is the length of all needed physical CUs. 
    \end{itemize}    
\end{itemize}

Since $N_{ori}$ actually represents the compressed size of user-requested data, we estimate it using a typical compression ratio of EBS production environments (50.1\%). 
Additionally, when {\compressunit} > 16 and $D_{new} \geq$ 16, two read operations are required, incurring additional overhead from metadata operations.
In {\org}'s production environment, the extra overhead from these dual reads is approximately equivalent to reading two additional blocks. Therefore, we account for this overhead by increasing $H_{size}$ by two blocks when estimating the performance overhead of the {\compressunit} alignment.

{\org} Sysadmins provide us 20 representative {\ebs} clusters' performance data, including metrics for $V_{disk}$, $V_{net}$, and $V_{dec}$.
Due to confidentiality concerns, we only show the normalized values of these metrics in \autoref{tab:performance-metrics}.
To show the impact of {\compressunit} alignment under different system loads, we use three sets of metrics to represent normal load (P50), high load (P90), and extreme load (P99) in \autoref{tab:performance-metrics}.

\begin{table}[t]
  \centering
  \caption{\textbf{Performance metrics of {\org}'s production environment.} $V_{disk} + V_{net}$ is the ratio of disk+network I/O time per block to end-to-end time. $V_{dec}$ is the ratio of decompression time per block to the end-to-end time.}
  \vspace{-5pt}
  \resizebox{0.58\linewidth}{!}{
    \begin{tabular}{cccc}
      \toprule
      \textbf{Metric} & \textbf{P50} & \textbf{P90} & \textbf{P99} \\ 
      \midrule
      $V_{disk} + V_{net} (\%) $ & 1.49 & 1.10 & 0.797 \\ 
      $V_{dec}$ (\%) & 2.50 & 2.13 & 1.92 \\ 
      \bottomrule
    \end{tabular}
  }
  \label{tab:performance-metrics}
  \vspace{-5pt}
\end{table}

We simulate the latency overhead of user read requests introduced by the adaptive compression scheme for 1.8\,k VDs using the above metrics and formulas.
As shown in \autoref{tab:cu_overhead}, the adaptive compression scheme only increases read latency by 0.477--2.60\% on average, which is acceptable.  

\begin{table}[t]
  \centering
  \caption{\textbf{Average read latency increased ratio.}}
  \label{tab:cu_overhead}
  \vspace{-5pt}
  \resizebox{0.78\linewidth}{!}{
    \begin{tabular}{lccc}
    \toprule
    \textbf{Workload} & \textbf{P50 (\%)} & \textbf{P95 (\%)} & \textbf{P99 (\%)} \\
    \midrule
    Elasticsearch & 2.31 & 1.83 & 1.43 \\
    BigData       & 2.60 & 2.07 & 1.59 \\
    Container     & 0.657 & 0.524 & 0.412 \\
    MongoDB       & 1.82 & 1.45 & 1.13 \\
    Message Queue            & 1.45 & 1.16 & 0.903 \\
    MySQL         & 0.769 & 0.610 & 0.477 \\
    Redis         & 1.92 & 1.53 & 1.19 \\
    WebApp        & 2.29 & 1.82 & 1.41 \\
    \bottomrule
    \end{tabular}
  }
  \vspace{-5pt}
\end{table}

%% file: appendix/split_proof.tex
\section{Proof about Leaf Overflow}
\label{sec:split-proof}

\begin{figure}[t]
  \centering
  \includegraphics[width=0.48\textwidth]{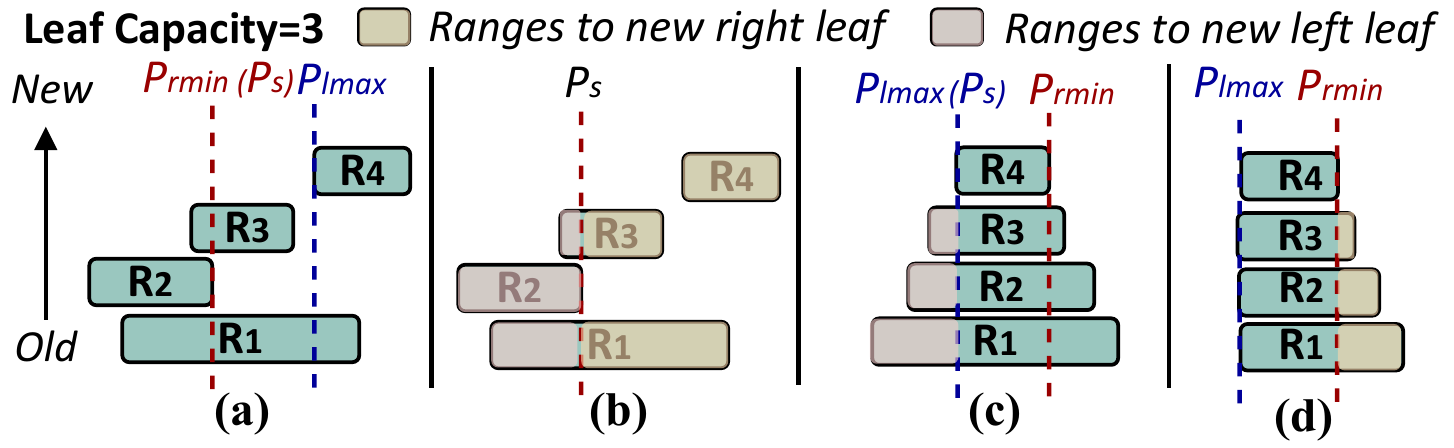}
  \vspace{-20pt}
  \caption{\textbf{Leaf overflow cases.}}
  \label{fig:split_overflow}
  \vspace{-5pt}
\end{figure}

Let the leaf capacity be $N$, the leaf and right bounds of the leaf's ranges and the new range to be inserted are $P_{l1}, \ldots, P_{l(N + 1)}$ and $P_{r1}, \ldots, P_{r(N + 1)}$, respectively;
$P_{lmax}$ and $P_{rmin}$ are the maximum value of $P_{li}$ and the minimum value of $P_{ri}$, respectively;
the selected split point is $P_s$.

\noindent
\myparagraph{Split regulation.}
Entries with range left bounds $\geq$ $P_s$ are moved to the new right leaf,
while entries across the $P_s$ 
are divided and stored in both leaves.
To minimize the range fragmentation,
when a range's left bound or right bound is equal to $P_s$, it is assigned to the new right leaf or left leaf, respectively.
In this case, if the left bound equals the right bound, the range is assigned to the new right leaf.

\noindent
\myparagraph{Theorem.} 
The split point selection in Section 5.3 guarantees:
\begin{itemize}[leftmargin=*, noitemsep, topsep=0pt]
  \item If $P_{lmax} > P_{rmin}$, no overflow occurs;
  \item If $P_{lmax} \leq P_{rmin}$, at least one new leaf will not overflow; if the other new leaf overflows, it will always be addressed by splitting it again (i.e., second split).
\end{itemize}

\paragraph{Proof.}
As the split point $P_s$ is one of the median points of all boundaries (i.e., $\{P_{li},P_{ri}\mid i=1,\dots,N+1\}$), it satisfies: $N + 1$ boundaries $\leq P_s$ and $N + 1$ boundaries $\geq P_s$.
Given that $P_{li} \leq P_{ri}$ for each range $i$, $P_s$ must satisfy one of the following two cases to ensure it is one of the median points:
\begin{itemize}[leftmargin=*, noitemsep, topsep=0pt]
  \item (Case 1: $P_{lmax} > P_{rmin}$) There exists at least one left bound $\geq P_s$ (i.e., $P_{lmax} \geq P_s$), and to satisfy $N + 1$ bounds $\leq P_s$, there must also exists a right bound $\leq P_s$ (i.e., $P_{rmin} \leq P_s$). 
  \item (Case 2: $P_{lmax} \leq P_{rmin}$) All left bounds $\leq P_s \leq$ all right bounds, in this case, $P_s$ is either $P_{lmax}$ or $P_{rmin}$. 
\end{itemize}

For Case 1 (e.g., \autoref{fig:split_overflow}(a)), there exists a range's left bound $\geq P_s$, 
and this range will be fully assigned to the new right leaf ($R_4$ in \autoref{fig:split_overflow}(b)).
Thus, the new left leaf has at most $N$ entrie, and it will not overflow.
At the same time, there exists a range's right bound $\leq P_s$ (i.e., $P_{rmin} \leq P_s$) and its left and right bounds cannot equal $P_s$ simultaneously 
because of $P_{rmin} \leq P_s \leq P_{lmax}$ and $P_{lmax} \neq P_{rmin}$. 
Thus, this range will be fully assigned to the new left leaf ($R_2$ in \autoref{fig:split_overflow}(b)) and the new right leaf will not overflow.
In summary, \emph{when $P_{lmax} > P_{rmin}$,  $P_s$ ensures both new leaves won't overflow.}

For Case 2, if $P_s = P_{lmax}$ (\autoref{fig:split_overflow}(c)), then the range with left bound $P_{lmax}$ will be fully assigned to the new right leaf ($R_{4}$ in \autoref{fig:split_overflow}(c)) and the new left leaf will not overflow.
Similarly, if $P_s = P_{rmin}$ (\autoref{fig:split_overflow}(d)), then the range with right bound $P_{rmin}$ will be fully assigned to the new left leaf ($R_{4}$ in \autoref{fig:split_overflow}(d)) and this leaf will not overflow.
In summary, \emph{when $P_{lmax} \leq P_{rmin}$, $P_s$ ensures one new leaf won't overflow.}

\begin{figure}[t]
  \centering
  \includegraphics[width=\linewidth]{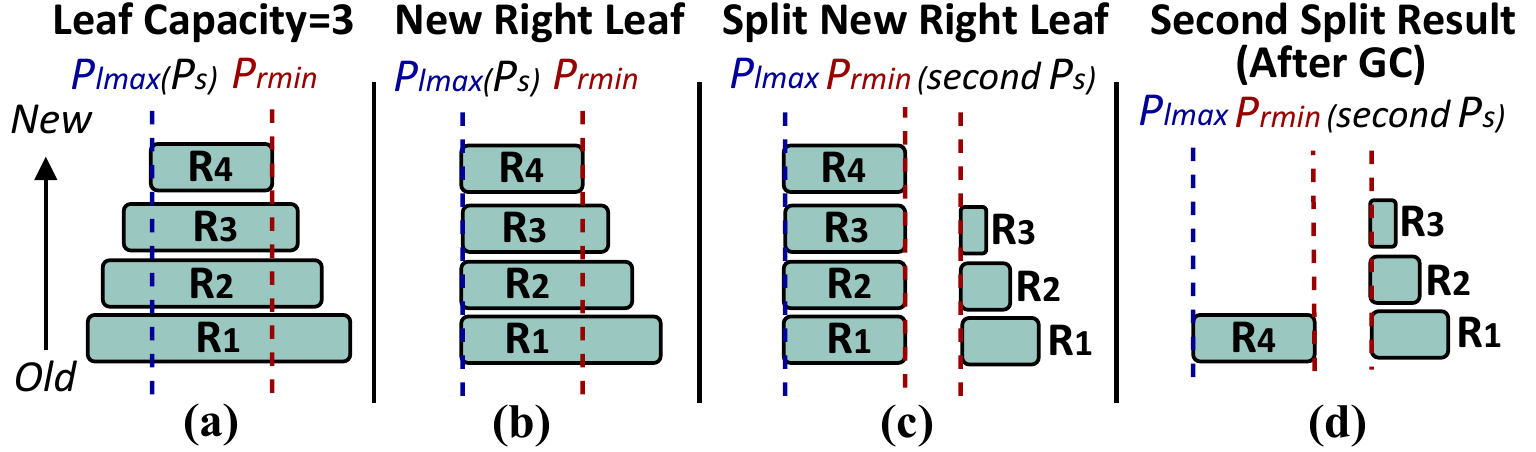}
  \vspace{-20pt}
  \caption{\textbf{Case triggering the second split.}}%
  \label{fig:split_proof}
  \vspace{-10pt}
\end{figure}

Then we prove that if the other new leaf overflows, the overflow will be resolved by a second split.
As shown in \autoref{fig:split_proof}(a),
we consider the case where $P_s = P_{lmax}$ (the case where $P_s = P_{rmin}$ is similar).
At this point, the new right leaf has $N + 1$ entries (N is the leaf capacity), and all entries' left bound in this leaf is $P_{lmax}$. 
Thus, as shown in \autoref{fig:split_proof}(b), the new right leaf will overflow (i.e., GC cannot remove any entry) only if the newer entry's right bound is smaller.
In this case, we perform another split on the new right leaf using its $P_{rmin}$ as $P_s$ according to the split point selection strategy in Section 5.3,
As shown in \autoref{fig:split_proof}(c) and \autoref{fig:split_proof}(d),
\emph{after the second split, no leaf will overflow} as:
(1) The new left leaf generated by the second split will have a range space between $P_{lmax}$ and $P_{rmin}$; after GC, only one entry will remain in this leaf.
(2) The new right leaf generated by the second split will have at most $N$ entries since at least one entry is fully assigned to the resulting new left leaf.

%% file: appendix/more_design.tex
\section{Additional Design and Implementation}
\label{sec:more_design}

\subsection{Customized optimizations for EBS}

In this section, we first introduce the value format of {\sys} when integrated into {\org}'s {\ebs}.
Then, given that 
\emph{{\org}'s {\ebs} is a single-threaded system with extremely high memory-efficiency requirements},
we customize the following two optimizations for {\ebs}:
proposing a two-level nested tree to improve performance and save memory,
and using a {\secondary} as a supplement to further save memory for value storage.
Based on the {\secondary}, we provide implementations of \codeword{DivideValue} and \codeword{MergeRange} functions for {\org}'s {\ebs}, which also serve as the default implementations in {\sys}.

\subsubsection{Value Format for {\org}'s {\ebs}}
\label{sub:value_format}

For {\org}'s {\ebs}, the value in {\sys} records the {\datafile} ID, and the offset and length in the {\datafile} for each {\seqwritestream}.
With the adaptive compression scheme (\autoref{sub:cu-align}), {\seqwritestream}s shorter than 4 blocks share a 4-block {\compressunit}.
To handle this, {\sys}'s value includes an additional 2-bit field to indicate the {\seqwritestream}'s offset (\emph{CU\textsubscript{off}}) within the {\compressunit}.
If \emph{CU\textsubscript{off}} $\neq$ 0, the recorded offset is the {\compressunit}'s start offset in the {\datafile} rather than the offset of the {\seqwritestream}.
The actual offset of the {\seqwritestream} needs to be calculated after decompression by adding $CU_{off} \times BlockSize$.

In the implementation, the value requires 10 bytes in total, saving 2 bytes compared to the {\ebs}-index (i.e., {\offsettable}\,+\,{\indexmap}).
This is because {\sys} unifies the indexing granularity to {\seqwritestream}, avoiding separately recording the {\compressunit} location and offset within the {\compressunit} in two indexes.

\begin{figure}[t]
  \centering
  \includegraphics[width=\linewidth]{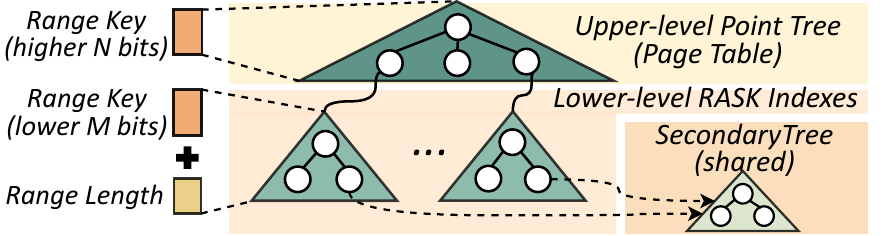}
  \vspace{-17pt}
  \caption{\textbf{Two-level nested tree.}}%
  \label{fig:arch}
  \vspace{-12pt}
\end{figure}

\subsubsection{Two-level Nested Tree}
{\plstree} needs key comparisons during internal node traversal 
due to the multi-entry leaf, sacrificing ART's efficient prefix matching.
We find that in {\ebs}, range lengths are typically much smaller than the key space (i.e., the maximum range length is only 128 due to the {\segmentcache}).
Thus, instead of using a single {\plstree} index for the entire key space, 
we use the two-level nested tree (\autoref{fig:arch}): we partition the key space into segments, which are indexed by an upper-level {\pask} tree (e.g., trie) for fast prefix matching (N bits);
each segment is managed by a {\plstree} index that indexes the remaining M bits of the range key and range length. 
Given that block devices use a continuous address space (0 to the highest address), {\sys} uses a two-level page-table as the upper-level tree to ensure high performance while controlling memory overhead.
All {\plstree} indexes share a {\secondary} (\autoref{sub:secondary}) to further save memory.
This optimization can further improve throughput by 16.7\% and reduce memory footprint by 10.9\%.

\subsubsection{{\secondary} as Supplement}
\label{sub:secondary}

\begin{figure}[t]
  \centering
  \includegraphics[width=0.95\linewidth]{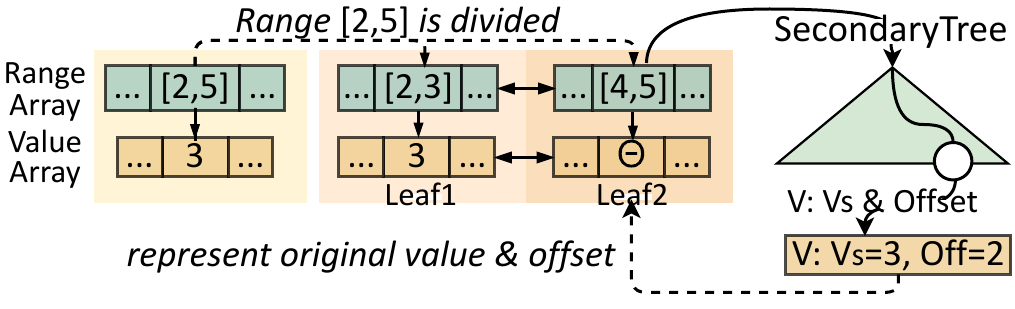}
  \vspace{-7pt}
  \caption{\textbf{{\secondary} example.} {\secondary}'s key is {\plstree} ID\,\&\,fragmented ranges.}%
  \label{fig:secondary}
  \vspace{-10pt}
\end{figure}

For {\ebs}, {\sys}'s value needs to record the offset and length of a {\compressunit} (i.e., {\seqwritestream}) in the {\datafile}.
Since the {\compressunit} cannot be physically divided in DFS when the value needs to be divided, all fragmented ranges\footnote{Fragmented ranges are subranges (except the first) after a range is divided.} must record their offsets (\emph{off}) from the original range's start 
and the original value ($V_{s}$) 
for correct value retrieval.
To avoid adding a 4-byte offset field to every {\plstree} entry,
as shown in \autoref{fig:secondary},
{\sys} uses a {\secondary}\,(an independent B+tree) to 
map fragmented ranges to their $V_{s}$ and \emph{off}.
For the fragmented range, an indicator\,($\Theta$)\footnote{Indicator is a value (all 1s for {\org} EBS) unused by regular value.} is stored in {\plstree}'s entry as its value.
During reads, if $\Theta$ is found, 
we use the current {\plstree} index's ID and the fragmented range as the key to retrieve the $V_{s}$ and \emph{off} from {\secondary}.

\begin{figure}
  \centering
  \includegraphics[width=\linewidth]{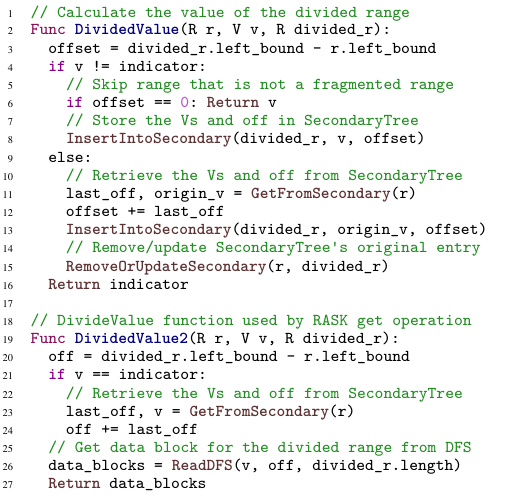}
	\caption{\textbf{Pseudocode of {\sys}'s DivideValue function for {\org}'s EBS.}}
	\label{fig:code:ebs}
\end{figure}

\begin{figure}
  \centering
  \includegraphics[width=\linewidth]{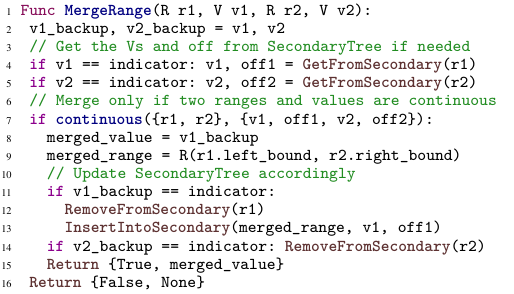}
	\caption{\textbf{Pseudocode of {\sys}'s MergeRange function for {\org}'s EBS.}}
	\label{fig:code:ebs2}
\end{figure}

In the implementation, we develop the user-provided functions (i.e., \codeword{DivideValue} and \codeword{MergeRange}) for EBS following the logic in \autoref{fig:code:ebs} and \autoref{fig:code:ebs2}.
These implementations ensure that the {\secondary} can record and retrieve the original value and offset for fragmented ranges correctly.
Notably, this implementation is independent of the value format, focusing only on the relationship between the divided range and the original range.
Thus, we use this implementation as the default for {\sys}'s \codeword{DivideValue} and \codeword{MergeRange} functions.
Users can customize more efficient implementations for other scenarios with different semantics.

In {\sys}'s \codeword{Get} operation,
the retrieved value may only correspond to a portion of the leaf range or be an indicator.
For example, when searching for [3,\,5] within leaf1 and leaf2 in \autoref{fig:secondary}, we only require a portion of the value from leaf1's entry [2, 3], and the value corresponding to leaf2's [4, 5] is an indicator.
As described in Section 6.1, {\sys} uses the \codeword{DivideValue} function to obtain the value corresponding to the target subrange.
However, the implementation of {\org}'s {\ebs} \codeword{DivideValue} function involves modifications to the {\secondary}.
Therefore, for {\org}'s {\ebs}, the \codeword{Get} operation actually uses another implementation of \codeword{DivideValue} (\codeword{DividedValue2} in \autoref{fig:code:ebs}) to retrieve the value for the subrange.
Specifically, for each value returned,
if it is an indicator, we retrieve the original value and offset from {\secondary}.
The original value corresponds to the location of the {\seqwritestream} in DFS (\autoref{sub:value_format}), while the offset and the subrange information can be used to calculate the position and length of the subrange within the {\seqwritestream}.
Then we read data from DFS according to the original value, decompress it, and return the data block corresponding to the needed subrange to the user.

The evaluation shows that the {\secondary} only occupies 0.158\% of {\sys}'s total memory consumption.
Its low memory footprint highlights its effectiveness in memory savings, 
reducing overall memory usage by 22.0\% compared to storing offsets in {\plstree}'s leaves,
despite a slight performance drop due to indirect access (3.42\%).

\subsubsection{Fault Tolerance Support}

In {\org}'s {\ebs}, to achieve high availability, the EBS-index must be periodically persisted to the DFS.
When using RASK to replace the EBS-index, we periodically flush the latest state of RASK to the DFS in the background.
This process is lightweight and efficient.
During recovery, we only need to load the latest RASK snapshot from the DFS. 
Any updates not yet flushed can be recovered by replaying the {\journal}.

\subsection{Range space-aware SMOs}
\label{sub:more_smo}

Splits may cause a too small range space for a leaf (i.e., range space is smaller than the leaf's capacity).
In this case, the leaf will never be fully filled with valid entries.
Although our split design (Section 5.3) does not address this issue explicitly, its median-based split point selection ensures this issue rarely occurs across various workloads (\textasciitilde2.1\%).
In this section, we  present two alternative split and merge/resplit strategies that explicitly take range space into account to address this issue.

\paragraph{Range space-aware split point selection.}
When taking the leaf's range space into account, the split point selection strategy in {\sys} considers two requirements.
\emph{R1}: It should not intersect with any range to avoid range fragmentation.
\emph{R2}: Each new leaf's {\representspace} should meet the theoretical minimum (i.e., the leaf's capacity $N$) to avoid space waste.

Since \emph{R1} is the same as the original split point selection strategy (Section 5.3),
we use the same set of split point candidates as the original strategy.
To meet \emph{R2}, we only consider split point candidates that ensure the new leaf's {\representspace}s is at least $N$.
Notably, 
it is impossible to satisfy \emph{R2} for both new leaves if the original leaf's range space < 2$N$.
In such cases, we relax the requirement to first ensure the new left leaf's {\representspace}, leaving right leaf issues to merge/resplit. 

When the split point selection causes leaf overflow, we split the overflowed leaf again as described in Section 5.3.

\paragraph{Range Space-aware merge and resplit.}
We further employ merge and resplit to reduce occurrences of too small leaf range space.
Apart from the trigger conditions described in Section 5.4, we also perform the merge/resplit when the leaf's range space falls below the theoretical minimum.
Specifically, we check the leaf's range space after inserting a new entry or during a background thread's periodic scan.
If the leaf's range space is smaller than the theoretical minimum, we execute the merge/resplit logic (Section 5.4) for this leaf and its right neighbor.
We perform merge/resplit with the right neighbor rather than the left because the split process prioritizes ensuring the left neighbor's range space is at least $N$.
Therefore, the left neighbor is less likely to have range space issues, and merging with it would yield less benefit.